\documentclass{amsart}
\usepackage{amssymb}
\usepackage{amsfonts}

\newtheorem{theorem}{Theorem}
\theoremstyle{plain}
\newtheorem{acknowledgement}{Acknowledgement}

\newtheorem{lemma}{Lemma}

\newtheorem{proposition}{Proposition}
\newtheorem{remark}{Remark}

\numberwithin{equation}{section}
\input{tcilatex}

\begin{document}
\title[Discrete dynamics of complex bodies]{Discrete dynamics of complex
bodies with substructural dissipation: variational integrators and
convergence}
\author{Matteo Focardi*}
\address{*Dipartimento di Matematica "U. Dini", Universit\`a di Firenze,
viale Morgagni 67/A, I-50139 Firenze (Italy)}
\email{focardi@math.unifi.it}
\author{Paolo Maria Mariano$^\dag$}
\address{$^\dag$DICeA, Universit\`a di Firenze, via Santa Marta 3, I-50139
Firenze (Italy)}
\email{paolo.mariano@unifi.it, paolo.mariano@math.unifi.it}
\date{\today}

\begin{abstract}
For the linearized setting of the dynamics of complex bodies we construct
variational integrators and prove their convergence by making use of BV
estimates on the rate fields. We allow for peculiar substructural inertia
and internal dissipation, all accounted for by a d'Alembert-Lagrange-type
principle.
\end{abstract}

\maketitle


\section{Introduction}

A \emph{variational integrator} is any recursive rule that allows one to
calculate discrete trajectories from initial data and coincides with the
discrete Euler-Lagrange equation of some discrete (or discretized)
Lagrangian. It is a discrete scheme for evaluating (numerically) the
dynamics described by some Hamilton or d'Alembert-Lagrange principle. A
detailed description of the essential properties of variational integrators
is in \cite{MW} (see also \cite{MPS}).

In particular, when the potential appearing in the Lagrangian functional can
be additively subdivided in subsystems, e.g. distinct groups of particles or
finite elements in space in case of continuous problems, it is possible to
select asynchronous time discretizations. Asynchronous variational
integrators (AVIs) proposed in \cite{LMOW1} and \cite{LMOW2} then follow.
They preserve the symplectic structure of the original continuum system.
Thus, for non-constant time steps the energy evaluated numerically
oscillates around its average value.

Although problems in evaluating individual trajectories are the same as for
other traditional alghoritms, AVIs perform better in determining
time-averaged quantities. Artificial numerical damping is not introduced.
For a desired error value the computational cost is lesser than the one of
other methods. AVIs display analogies with subciclyc methods (see \cite{Bel}
and \cite{BM}) and other methods in molecular dynamics (appropriate
comparisons are in \cite{LMOW2}).

AVIs have been introduced in \cite{LMOW1} for the dynamics of elastic simple
bodies. Their convergence in time has been analyzed variously: A repeated
use of Gronwall inequality on the discrete Euler-Lagrange equations at a
specific node is the basic ingredient of the technique introduced in \cite%
{LMOW2} for potentials with uniformly bounded second derivative. $\Gamma $%
-convergence has been exploited for the zero-dimensional oscillator in \cite%
{MO}, some technical hypotheses removed in \cite{MM}. For the linear
elastodynamics of simple bodies, the convergence has been also proven in 
\cite{FM} along a path based on the direct exploitation of the action
functional.

Here we start the analysis of variational integrators (in the synchronous
and asynchronous versions) for the linear dynamics of complex bodies, those
bodies for which the material substructure prominently influence the gross
behavior. We consider cases in which the material substructures display
peculiar inertia (additional to the macroscopic one) and internal
dissipation. The dynamics of such bodies is governed by a
d'Alembert-Lagrange-type principle.

We do no treat any specific model of complex materials. We develop our
analyses in the general model-building framework of the mechanics of complex
bodies (\cite{C89}, \cite{M02}). In this way we do not specify the type of
the substructure under analysis. We presume only that its essential
geometrical features are described by an element of an abstract finite
dimensional manifold $\mathcal{M}$, the so-called manifold of substructural
shapes. We construct first the linearized version of the natural non-linear
theory. To this aim we need the isometric emebedding of $\mathcal{M}$ in a
linear space. Although such embedding always exists, it is not unique, so
its choice is a constitutive part of the modeling procedure. After
establishing the structure of the linear theory, we construct appropriate
variational integrators and we prove the convergence of the discrete
approximation in time to the continuous counterpart, the convergence in
space being assured by standard theorems.

The technique that we use here is new: it is based essentially on \emph{BV}
estimates on the time-rates of the fields involved.

Convergence in the asynchronous case is proven only for substructures
admitting quadratic peculiar independent kinetic energy. The expression of
the substructural kinetic energy is accepted in fully generality in proving
convergence of synchronous variational integrators.\ Our work prepare the
way to a number of numerical experiments in various classes of complex
bodies.

\textbf{Some notations}. We use the standard symbols $|\cdot |$ for the
euclidean norm, and both $\cdot $ and $\langle \cdot ,\cdot \rangle $ for
the scalar product in some Euclidean space $\mathbb{R}^{j}$, $j\in \mathbb{N}
$. We will not explicitly indicate the dependence on the dimension $j$, for
the sake of the conciseness of the notation. The symbol $\#$ denotes the
cardinality: $\#\mathcal{T}$ is then the cardinality of the set $\mathcal{T}$%
. By $\partial _{z}$ we indicate the partial derivative with respect to $z$.
We use standard notations for Lebesgue spaces, Sobolev spaces, and the space
of functions with bounded variation ($BV$ in short). The symbol $\mathcal{H}%
^{d-1}$ denotes the $(d-1)$-dimensional Hausdorff measure in $\mathbb{R}^{d}$%
. In some computations we find inequalities involving constants which depend
on the data and on the space dimensions, but are always independent of the
set of discrete instants. Since it is not essential to distinguish from one
specific constant to another, we indicate all of them by the same letter $c$%
, leaving understood that $c$ may change from one inequality to another.

\section{Commentary to the dynamics of complex bodies}

\label{complxdyn}

In a primitive approach, a body is considered as an abstract set $\mathfrak{B%
}$ the elements of which are called \emph{material elements}. Each element
is considered as the smallest piece of matter characterizing the nature of
the material constituting the body under examination, that is a patch of
matter made of entangled molecules or the characteristic cell of some atomic
lattice. The representation of this set is a model of the morphology of the
body. It cannot prescind from the selection of the placements of the body in
the ambient space $\mathbb{R}^{d}$, placements that are selected by means of
one-to-one maps $f:\mathfrak{B}\longrightarrow \mathbb{R}^{d}$. The generic $%
\mathcal{B}:=f\left( \mathfrak{B}\right) $ represents the `gross' shape of
the body and is assumed to be a bounded domain with boundary $\partial 
\mathcal{B}$ of finite $\left( d-1\right) $-dimensional measure, a boundary
where the outward unit normal $n$ is defined to within a finite number of
corners and edges. When substructural complexity arises, description of the
structure inside material elements is necessary. The representation of $%
\mathfrak{B}$ is then defined by maps $\mathfrak{\varkappa :B}\rightarrow 
\mathcal{M}$ assigning to each material element a descriptor of the peculiar
features of its inner morphology (called for this reason a morphological
descriptor), with $\mathcal{M}$ a finite-dimensional differentiable manifold.

The dynamics of a complex body involves then the description of subsequent
changes in gross and substructural shapes.

Convenience suggests to select a reference place $\mathcal{B}_{\ast
}:=f_{\ast }\left( \mathfrak{B}\right) $ that can be in principle occupied
by the body in $\mathbb{R}^{d}$ and to obtain from it new places by means of
sufficiently smooth bijective maps (transplacements)%
\begin{equation*}
x\longmapsto y=y\left( x\right) =\left( f\circ f_{\ast }^{-1}\right) \left(
x\right) \in \mathbb{R}^{d},\text{ \ \ }x\in \mathcal{B}_{\ast },
\end{equation*}%
such that (\emph{i}) $\mathcal{B}=y\left( \mathcal{B}_{\ast }\right) $\ is
as regular as $\mathcal{B}_{\ast }$ and (\emph{ii}) the \emph{gradient of
deformation} 
\begin{equation*}
F=\nabla y\in Hom\left( T_{x}\mathcal{B}_{\ast },T_{y}\mathcal{B}\right)
\end{equation*}%
has positive determinant at $x\in \mathcal{B}_{\ast }$. A time-parametrized
family of deformations 
\begin{equation*}
\left( x\mathbf{,}t\right) \longmapsto y=y\left( x\mathbf{,}t\right) \in 
\mathbb{R}^{d},\text{ \ \ }\left( x\mathbf{,}t\right) \in \mathcal{B}_{\ast
}\times \left[ 0,\bar{t}\right] ,
\end{equation*}%
is the motion at gross scale. Sufficient smoothness in time is presumed so
that the velocity $\dot{y}=\frac{dy}{dt}\left( x\mathbf{,}t\right) \in
T_{y\left( x,t\right) }\mathcal{B}\simeq \mathbb{R}^{d}$ can be defined as a
field over $\mathcal{B}_{\ast }$.

Differentiable maps of the type%
\begin{equation*}
x\longmapsto \nu :=\left( \varkappa \circ f_{\ast }^{-1}\right) \left(
x\right) \in \mathcal{M},\text{ \ \ }x\in \mathcal{B}_{\ast },
\end{equation*}%
define fields of morphological descriptors. Then, a substructural `motion'
is then given by%
\begin{equation*}
\left( x\mathbf{,}t\right) \longmapsto \nu =\nu \left( x\mathbf{,}t\right)
\in \mathcal{M},\text{ \ \ }\left( x\mathbf{,}t\right) \in \mathcal{B}_{\ast
}\times \left[ 0,\bar{t}\right] .
\end{equation*}%
Assuming sufficient smoothness in time, the rate of change of the
substructural morphology in the referential description is $\dot{\nu}:=\frac{%
\partial \nu }{\partial t}\left( x\mathbf{,}t\right) $. At each $x$ the
spatial derivative of $x\longmapsto \nu $ is indicated by $N\in Hom\left(
T_{x}\mathcal{B}_{\ast },T_{\nu }\mathcal{M}\right) $.

In a conservative setting the mechanics of a complex body is governed by
Hamilton principle of least action with Lagrangian density 
\begin{equation*}
\mathcal{L}=\frac{1}{2}\rho \left\vert \dot{y}\right\vert ^{2}+\chi \left(
\nu ,\dot{\nu}\right) -\rho e\left( x\mathbf{,}F,\nu ,N\right) -\rho w\left(
y\mathbf{,}\nu \right) ,
\end{equation*}%
assumed to be differentiable with respect to all its entries. $\rho $ is the
referential mass density (conserved during the motion), $e$ the elastic
potential and $w$ the potential of external actions, all per unit mass. The
term $\chi \left( \nu ,\dot{\nu}\right) $ is a real-valued function
vanishing when $\dot{\nu}=0$, it is of degree 2 in $\dot{\nu}$ and is such
that the second derivative $\partial _{\dot{\nu}\dot{\nu}}^{2}\chi $ exists
and is positive definite. Moreover, $\sup_{\dot{\nu}}\left( \partial _{%
\mathbf{\dot{\nu}}}\chi \cdot \dot{\nu}-\chi \right) $ is attained (as a
maximum) at a unique point in the relevant $T_{\nu }\mathcal{M}$. It
coincides with the substructural kinetic energy when peculiar independent
inertia pertains to substructural changes (see remarks in \cite{CG}, \cite%
{M07}).

The derivatives of $\mathcal{L}$ with respect to $F$ and $N$ represent
contact interactions, namely standard and microscopic stresses,
respectively. The derivatives of $\mathcal{L}$ with respect to $y$ and $\nu $
are standard and substructural bulk actions, the latter being divided
additively in external fields acting on the substructure and internal
self-actions.

Dissipative effects may occur. For the sake of simplicity it is assumed here
that dissipation occurs only within each material element and is accounted
for by the local self-actions. A concrete example is given by quasicrystals
(see \cite{M05}, \cite{M07}).

A d'Alembert-Lagrange-type type principle then substitutes the Hamiltonian
one. One then requires that%
\begin{equation}
\delta \int_{\mathcal{B}_{\ast }\times \lbrack 0,\bar{t}]}\mathcal{L}\text{ }%
dx\wedge dt+\int_{\mathcal{B}_{\ast }\times \lbrack 0,\bar{t}]}z^{v}\cdot
\phi \text{ }dx\wedge dt=0,  \label{dAL}
\end{equation}%
where $\delta $ indicates the first variation, $z^{v}=z^{v}\left( F,\nu ,%
\dot{\nu},N\right) \in T_{\nu }^{\ast }\mathcal{M}$ represents the
dissipative part of the self-actions inside the generic material element and 
$(x,t)\longmapsto \phi :=\phi \left( x,t\right) \in T_{\nu \left( x,t\right)
}\mathcal{M}$ is an arbitrary smooth field with compact support.

To define variations, for $\varepsilon \in \left[ 0,1\right] $ one considers
smooth maps%
\begin{equation*}
\varepsilon \longmapsto y_{\varepsilon }:=y+\varepsilon \varphi ,\quad \quad
\varepsilon \longmapsto \nu _{\varepsilon },
\end{equation*}%
with $\frac{d\nu _{\varepsilon }}{d\varepsilon }\left\vert _{\varepsilon
=0}\right. \left( x,t\right) =\phi \left( x,t\right) $, $(x,t)\longmapsto
\varphi :=\varphi \left( x,t\right) $ a smooth $\mathbb{R}^{d}-$valued map.
The variation $\delta $ along $\left( y,\nu \right) $ has then the meaning
of the derivative $\frac{d}{d\varepsilon }$ of $\mathcal{L}$ evaluated at $%
\varepsilon =0$. It is also assumed that the self-action $z^{v}$ is \emph{%
intrinsically dissipative}, that is%
\begin{equation*}
z^{v}\cdot \dot{\nu}\geq 0,
\end{equation*}%
for any choice of $\dot{\nu}$. Then $z^{v}$ may be of the type 
\begin{equation*}
z^{v}=\eta \dot{\nu}^{\flat },
\end{equation*}%
with $\dot{\nu}^{\flat }$ the covector associated with $\dot{\nu}$ and $\eta 
$ a positive diffusion coefficient.

When the fields $x\longmapsto P:=-\rho \partial _{F}\mathcal{L}$ and $%
x\longmapsto \mathcal{S}:=\rho \partial _{N}e$ are of class $C^{1}\left( 
\mathcal{B}_{\ast }\right) \cap $ $C^{0}\left( \mathcal{\bar{B}}_{\ast
}\right) $, Euler-Lagrange equations are then%
\begin{equation*}
\rho \ddot{y}=DivP+b,
\end{equation*}%
\begin{equation*}
\overset{\cdot }{\overline{\partial _{\dot{\nu}}\chi }}-\partial _{\nu }\chi
=Div\mathcal{S}-z+\beta .
\end{equation*}%
$P$ is the Piola-Kirchhoff stress tensor, $b:=-\rho \partial _{y}w$ the
co-vector of body forces, $\mathcal{S}$ the microstress tensor in
referential configuration, $z:=\rho \partial _{\nu }e$ the self-action, $%
\beta :=-\rho \partial _{\nu }w$ external body action over the substructure.

By considering fields over the actual place $\mathcal{B}$, the balance
equations above become 
\begin{equation*}
\rho _{e}a=div\sigma +b_{e},
\end{equation*}%
\begin{equation*}
\overset{\cdot }{\overline{\partial _{\dot{\nu}_{e}}\chi }}-\partial _{\nu
_{e}}\chi =Div\mathcal{S}_{e}-z_{e}+\beta _{e},
\end{equation*}%
where%
\begin{equation*}
\sigma :=\left( \det F\right) ^{-1}PF^{\ast },\text{ \ \ \ \ }b_{e}:=\left(
\det F\right) ^{-1}b,
\end{equation*}%
\begin{equation*}
\mathcal{S}_{e}:=\left( \det F\right) ^{-1}\mathcal{S}F^{\ast },\text{ \ \ \
\ }z_{e}:=\left( \det F\right) ^{-1}z,\text{ \ \ \ }\beta _{e}:=\left( \det
F\right) ^{-1}\beta .
\end{equation*}

A requirement of objectivity involving the action of $SO\left( 3\right) $ on
both the ambient space $\mathbb{R}^{d}$\ and $\mathcal{M}$ for the density
of the elastic potential $e$ implies that 
\begin{equation*}
skwPF^{\ast }=\mathsf{e}\left( \mathcal{\bar{A}}^{\ast }z+\left( D\mathcal{%
\bar{A}}\right) ^{\ast }\mathcal{S}\right) ,
\end{equation*}%
where $\mathcal{\bar{A}=\bar{A}}\left( \nu \right) \in Hom\left( \mathbb{R}%
^{d},T_{\nu }\mathcal{M}\right) $, and $\mathsf{e}$ is Ricci's alternating
symbol (see \cite{C89}). Moreover, one may require the invariance of the
entire Lagrangian under the action of the group of diffeomorphisms of the
ambient space and the group of diffeomorphisms of $\mathcal{M}$ into itself.
Such invariance requirement opens a path that allows one to prove the
covariance of the balance laws (see \cite{M07}).

Boundary conditions can be assigned in terms of places and morphological
descriptors on parts $\partial \mathcal{B}_{\ast y}\ $and $\partial \mathcal{%
B}_{\ast \nu }$ of the boundary $\partial \mathcal{B}_{\ast }$. Tractions $%
\mathsf{t}:=Pn$ and $\varpi $ $:=\mathcal{S}n$, with $n$ the normal to the
boundary in all places in which it is defined, can be in principle assigned
on parts $\partial \mathcal{B}_{\ast \mathsf{t}}\ $and $\partial \mathcal{B}%
_{\ast \varpi }$ of the same boundary. Mixed boundary conditions require $%
\partial \mathcal{B}_{\ast y}\cap \partial \mathcal{B}_{\ast \mathsf{t}%
}=\varnothing $, $\partial \mathcal{B}_{\ast \nu }\cap \partial \mathcal{B}%
_{\ast \varpi }=\varnothing $ and $\overline{\partial \mathcal{B}_{\ast
y}\cup \partial \mathcal{B}_{\ast \mathsf{t}}}=\overline{\partial \mathcal{B}%
_{\ast \nu }\cup \partial \mathcal{B}_{\ast \varpi }}=\varnothing $.

A loading device able to prescribe at the boundary the substructural contact
interactions $\varpi $ is not know in imagination or elsewhere. The sole
natural boundary conditions in terms of substructural contact interactions
is then $\varpi =0$. Different is the case of Dirichlet data in terms of
morphological descriptors. Example is the case of quasicrystals in nematic
order: surfactants spread along the boundary allow to assign a precise
orientation to the rod-like molecules in nematic order.

Initial conditions are given by regular fields $\left( x,t_{0}\right)
\rightarrow (y_{0}(x,t_{0}),\nu _{0}(x,t_{0}))$ and $\left( x,t_{0}\right)
\rightarrow (\dot{y}_{0}(x,t_{0}),{\dot{\nu}}_{0}(x,t_{0}))$.

\section{Linearization}

\label{linear}

In the standard theory of elasticity, the space of transplacements can be
considered as an infinite dimensional manifold. Linearization is developed
over the tangent space of such a manifold. In such a procedure it is useful
to define the \emph{displacement field}%
\begin{equation*}
\left( x,t\right) \longmapsto u:=u\left( x,t\right) =y\left( x,t\right) -x,%
\text{ \ \ }\left( x\mathbf{,}t\right) \in \mathcal{B}_{\ast }\times \left[
0,\bar{t}\right] .
\end{equation*}%
The initial condition is then given by $\left( x,t_{0}\right) \rightarrow
u_{0}(x,t_{0})$, and the boundary condition is given along $\partial 
\mathcal{B}_{\ast u}=\partial \mathcal{B}_{\ast y}$.

The condition $\left\vert \nabla u\right\vert <<1$ defines the infinitesimal
deformation regime in which, essentially, $\sigma \approx P$, and the stress
may depend linearly on the infinitesimal strain $\varepsilon :=sym\nabla u$.

To linearize the multifield setting described in previous section, an
additional difficulty arises because $\mathcal{M}$ does not coincide in
general with a linear space isomorphic to $\mathbb{R}^{k}$\ for some $k$.
For this reason, the construction of a Sobolev space of maps taking values
on $\mathcal{M}$ requires the embedding of the manifold of substructural
shapes itself in a linear space. We assume that $\mathcal{M}$ is endowed
with a $C^{1}$ Riemannian structure and the relevant Levi-Civita parallel
transport. By Nash theorem an isometric embedding of $\mathcal{M}$\ in a
linear space is always available but it is not unique. The selection of an
embedding becomes then a constitutive ingredient of each special model. Once
the embedding of $\mathcal{M}$ has been chosen, one may consider the space $%
\mathcal{C}$ of pairs of maps $\left( y,\nu \right) $ can be considered as
an infinite dimensional differentiable manifold.

We maintain the assumption of infinitesimal deformation setting, a regime in
which we can `confuse' $\mathcal{B}_{\ast }$ with $\mathcal{B}$, in the
sense that if $\varphi $ indicates a vector field tangent to $y$, then $%
\varphi \approx u$ at any $x$ in $\mathcal{B}_{\ast }$. For this reason,
from now on we write $\mathcal{B}$ instead of $\mathcal{B}_{\ast }$\ for the
sake of conciseness. Moreover, we also confuse $\nu $ with $\nu _{e}=\nu
\circ y^{-1}$ for consequent obvious reasons.

We consider differentiable\ fields%
\begin{equation*}
\left( x,t\right) \longmapsto \varphi \left( x,t\right) \in T_{u}\mathbb{R}%
^{d},\text{ \ \ }\left( x,t\right) \in \mathcal{B}\times \left[ 0,\bar{t}%
\right] ,
\end{equation*}%
\begin{equation*}
\left( x,t\right) \longmapsto \phi \left( x,t\right) \in T_{\nu }\mathcal{M},%
\text{ \ \ }\left( x,t\right) \in \mathcal{B}\times \left[ 0,\bar{t}\right] ,
\end{equation*}%
such that $\varphi $ and $\phi $ vanish where Dirichlet data are imposed. We
continue to write $\nu $ and $\mathcal{M}$ as before the embedding. The
notation $L\left( A\right) \left( \bar{u},\varphi \right) \left( \bar{\nu}%
,\phi \right) $ indicates the linearization of $A$ about the pair of maps $%
\left( \bar{u}\mathbf{,}\bar{\nu}\right) $. A superposed bar denotes maps
calculated at $\left( \bar{u}\mathbf{,}\bar{\nu}\right) $. At each $x$ and $%
t $, the linearizations of the maps $x\longmapsto F$ and $\nu \longmapsto N$
about $\left( \bar{u}\mathbf{,}\bar{\nu}\right) $ are then given by%
\begin{equation*}
L\left( F\right) \left( \bar{u},\varphi \right) =\bar{F}+\nabla \varphi
\end{equation*}%
and%
\begin{equation*}
L\left( N\right) \left( \bar{\nu},\phi \right) =\overline{N}+\nabla \phi 
\mathbf{.}
\end{equation*}%
Previous formulas make sense because of pointwise parallel transport of $F$
over curves on $\mathbb{R}^{d}$, and of $N$ over $\mathcal{M}$.

The next step is to consider maps%
\begin{equation*}
\varsigma \longmapsto P\circ \varsigma ,\text{ \ \ }\varsigma \longmapsto
z\circ \varsigma ,\text{ \ \ }\varsigma \longmapsto \mathcal{S}\circ
\varsigma ,
\end{equation*}%
where $\varsigma =\left( F,\nu ,N\right) $, and to linearize them about $%
\left( \bar{u}\mathbf{,}\bar{\nu}\right) $. The infinitesimal deformation
setting implies $P\approx \sigma $, $z\approx z_{e}$, $\mathcal{S\approx S}%
_{e}$ and we continue to write $P$, $z$, $\mathcal{S}$\ in this sense. By
presuming sufficient smoothness, we then get%
\begin{eqnarray*}
&&L\left( P\right) \left( \bar{u},\varphi \right) \left( \bar{\nu},\phi
\right) =\bar{P}+\mathbf{L}^{\left( P\right) }\left( \nabla \varphi \mathbf{,%
}\phi \mathbf{,}\nabla \phi \right) , \\
&&L\left( z\right) \left( \bar{u},\varphi \right) \left( \bar{\nu},\phi
\right) =\bar{z}+\mathbf{L}^{\left( z\right) }\left( \nabla \varphi \mathbf{,%
}\phi \mathbf{,}\nabla \phi \right) , \\
&&L\left( \mathcal{S}\right) \left( \bar{u},\varphi \right) \left( \bar{\nu}%
,\phi \right) =\mathcal{\bar{S}}+\mathbf{L}^{\left( \mathcal{S}\right)
}\left( \nabla \varphi \mathbf{,}\phi \mathbf{,}\nabla \phi \right) ,
\end{eqnarray*}%
where the $\mathbf{L}^{\left( \cdot \right) }$'s are linear forms of their
entries. For the sake of simplicity we may assume that $\left( \bar{u}%
\mathbf{,}\bar{\nu}\right) $ are associated with a (so-called) natural state
in which $\bar{P}=0$, $\bar{z}=0$, $\mathcal{\bar{S}}=0$ (that is $\bar{%
\sigma}=0$, $\bar{z}_{e}=0$, $\mathcal{\bar{S}}_{e}=0$), so that%
\begin{eqnarray*}
&&L\left( P\right) \left( \bar{u},\varphi \right) \left( \bar{\nu},\phi
\right) =\mathbf{L}^{\left( P\right) }\left( \nabla \varphi \mathbf{,}\phi 
\mathbf{,}\nabla \phi \right) , \\
&&L\left( z\right) \left( \bar{u},\varphi \right) \left( \bar{\nu},\phi
\right) =\mathbf{L}^{\left( z\right) }\left( \nabla \varphi \mathbf{,}\phi 
\mathbf{,}\nabla \phi \right) , \\
&&L\left( \mathcal{S}\right) \left( \bar{u},\varphi \right) \left( \bar{\nu}%
,\phi \right) =\mathbf{L}^{\left( \mathcal{S}\right) }\left( \nabla \varphi 
\mathbf{,}\phi \mathbf{,}\nabla \phi \right) .
\end{eqnarray*}%
In this case, since the measures of interaction listed above are the partial
derivatives of $e$ with respect to $F$, $\nu $ and $N$, respectively, the
elastic energy is then a quadratic form in $\left( \nabla \varphi \mathbf{,}%
\phi \mathbf{,}\nabla \phi \right) $. By some abuse of notation, by taking
the common notation of standard linear elasticity, we then write%
\begin{equation*}
e=e\left( U\mathbf{,}\nu \mathbf{,}N\right)
\end{equation*}%
leaving understood the meaning of the entries, with $U=F-Id$ and $Id$ the
identity.

For the sake of simplicity and without reducing drastically the generality
of the physical cases covered by our treatment, the assumptions listed below
apply.

\begin{itemize}
\item[(A1)] The density of the elastic potential $e:\mathbb{R}^{d\times
d}\times \mathbb{R}^{k}\times \mathbb{R}^{k\times d}\rightarrow \lbrack
0,+\infty )$ is a \emph{continuous} and \emph{coercive quadratic form}.
There exist constants $\lambda ,\Lambda >0$ such that for every $(U,\nu ,N)$ 
\begin{equation*}
\lambda (|U|^{2}+|\nu |^{2}+|N|^{2})\leq e(U,\nu ,N)\leq \Lambda
(|U|^{2}+|\nu |^{2}+|N|^{2}).
\end{equation*}

\item[(A2)] The potential of external actions $w\in C^{1}({{\mathbb{R}}^{d}}%
\times {{\mathbb{R}}^{k}},[0,+\infty ))$ has \emph{quadratic growth}, i.e.
there exists $\Xi _{1}>0$ such that for every $(u,\nu )$ 
\begin{equation*}
|\nabla w(u,\nu )|\leq \Xi _{1}(1+|u|+|\nu |).
\end{equation*}

\item[(A3)] The traction $\mathsf{t}\in L^{2}(\partial \mathcal{B},\mathcal{H%
}^{d-1})$ is applied on a closed subset $\partial \mathcal{B}_{\mathsf{t}}$
of $\partial \mathcal{B}$. Moreover, a homogeneous boundary condition is
prescribed for the displacement $u$ on $\partial \mathcal{B}_{\mathsf{u}}:=%
\overline{\partial \mathcal{B}\setminus \partial \mathcal{B}_{\mathsf{t}}}$,
and for the morphological descriptor $\nu $ on a closed subset $\partial 
\mathcal{B}_{\nu }$ of $\partial \mathcal{B}$.
\end{itemize}

Such assumptions cover all special cases of complex bodies that we know in
such linearized setting (see \cite{MS} for a list of examples).

\section{Asynchronous Variational Integrators}

\label{AVIs}

The asynchronous variational integrator scheme necessary for the
d'Alembert-Lagrange-type principle \eqref{dAL} can be constructed easily by
following the guidelines of \cite{LMOW1} and \cite{LMOW2}. However, the
analysis of the convergence of AVIs in presence of an abstract form of the
substructural kinetic co-energy is hard. For this reason we restrict here
the attention to the case in which the kinetic co-energy $\chi $ has the
form 
\begin{equation*}
\chi (\nu ,{\dot{\nu}})=\frac{1}{2}\overline{\rho }|{\dot{\nu}}|^{2},
\end{equation*}%
where $\overline{\rho }$ is a material parameter. The general case is
treated later for synchronous variational integrators.

Of course the absence of dependence of $\chi $ on $\nu $ itself implies that 
$\mathcal{M}$ is a linear space per se and that an isometric parallelism is
selected over $\mathcal{M}$. The Levi-Civita connection resting on a
Riemaniann structure mentioned earlier is the case. Microcraked soft bodies
and incommensurate intergrowth compounds fall within this scheme (see \cite%
{M05} and \cite{MS} for relevant examples).

In this case, we then consider the linearized dynamics of a complex body
with zeroth-order dissipation governed by the d'Alembert-Lagrange-type
principle 
\begin{equation}
\delta \mathcal{A}(u,\nu )[\varphi ,\phi ]+\mathcal{D}(\dot{\nu})[\phi ]=0,
\label{ddALcont}
\end{equation}%
which is valid for arbitrary choices of test functions $\varphi $, $\phi $,
defined earlier. In (\ref{ddALcont}), $\mathcal{A}$ is defined by 
\begin{equation}
\mathcal{A}\left( \mathcal{B},\left[ t_{0},t_{f}\right] ;u,\nu \right)
:=\int_{t_{0}}^{t_{f}}\left( \int_{\mathcal{B}}\left( \frac{1}{2}\rho |\dot{u%
}|^{2}+\frac{1}{2}\overline{\rho }|{\dot{\nu}}|^{2}\right) dx-V\left( 
\mathcal{B},u,\nu ,t\right) \right) dt,  \label{action}
\end{equation}%
with%
\begin{equation}
V\left( \mathcal{B},u,\nu ,t\right) :=\int_{\mathcal{B}}\rho \left( e(\nabla
u,\nu ,\nabla \nu )+w(u,\nu )\right) dx-\int_{\partial \mathcal{B}_{\mathsf{t%
}}}\mathsf{t}\cdot u\,d\mathcal{H}^{d-1}  \label{potential}
\end{equation}%
and $\mathcal{D}$ is the dissipation given by%
\begin{equation}
\mathcal{D}(\dot{\nu})[\phi ]:=\eta \int_{\mathcal{B}\times I}{\dot{\nu}}%
\cdot \phi \,dx\wedge dt,  \label{Foc-dissipation}
\end{equation}
$\eta$ is a positive parameter.

\subsection{Spatial discretization}

\label{spacetime}

Discretization in space is obtained by means of standard finite elements
given by a tessellation $\mathcal{T}$ of $\mathcal{B}$, for instance a
regular triangulation chosen to be consistent with the partition of the
boundary $\partial \mathcal{B}$ into $\partial \mathcal{B}_{t}$ and $%
\partial \mathcal{B}_{u}$ (see, e.g., \cite{BLM}, \cite{DL}). By adopting a
triangulation we assume that $\mathcal{B}$ is a \emph{polyhedral domain}.
The integration nodes, the generic one being indicated by $a$, are selected
as the vertexes of the elements $K$ of the triangulation $\mathcal{T}$.

We consider the space $PA\left( \mathcal{T}\right) $ of continuos functions
which are linear polynomials on each $K\in \mathcal{T}$. With fixed a
bounded open interval $I=(t_{0},t_{f})\subseteq \mathbb{R}$, we are
interested in the vector subspace $\mathcal{V}_{u}$ of $PA\left( \mathcal{T}%
\right) \otimes H^{1}\left( I,{{\mathbb{R}}^{d}}\right) $ which includes all
the displacement mappings satisfying $u\left\vert _{\partial \mathcal{B}%
_{u}}\right. \equiv 0$. Any map $u\in \mathcal{V}_{u}$ has then the form 
\begin{equation}
u(x,t)=\sum_{a\in \mathcal{T}}\mathcal{N}_{a}(x)u_{a}(t),  \label{Shape}
\end{equation}%
where $\mathcal{N}_{a}\in PA\left( \mathcal{T}\right) $ is the nodal shape
function corresponding to node $a$ 
and $u_{a}(t)$ is the value of the displacement at $a$ at time $t$. Of
course we assume $u_{a}(t)\equiv 0$ if $a\in \partial \mathcal{B}_{u}$. The
null condition is not a restriction to our treatment.

With a slight abuse of notation, we denote by $u\left( t\right) $ the vector
in ${{\mathbb{R}}^{D}}$, with $D$ the number of degrees of freedom of all
nodal placements at time $t$, relative to the map $u\left( \cdot ,t\right) $%
. Specifically, $D=d\#{\mathcal{T}}$, with $\#{\mathcal{T}}$ the total
number of nodes in $\mathcal{T}$.

It is possible to check that there exists a constant $c_{\mathcal{T}}>0$,
depending only on the tessellation $\mathcal{T}$, such that 
\begin{equation}  \label{stimagrad}
\sup_{x\in\mathcal{B}}\left\vert \nabla u(\cdot,t) \right\vert \leq c_{%
\mathcal{T}}\left\vert u(t) \right\vert .
\end{equation}

When restricted to a generic finite element $K$, the map $u(x,t)$ in (\ref%
{Shape}) is indicated by $u_{K}(x,t)$. Analogously $u_{K}(t)$ indicates a
vector in $\mathbb{R}^{d\times (d+1)}$ having as components the
displacements of all nodes of the generic element $K$.

We perform an analogous spatial discretization for the field $\nu $. In this
case we have $\nu _{a}(t)\in {{\mathbb{R}}^{k}}$ for every $a\in {\mathcal{T}%
}$, $t\in I$ as a consequence of the embedding of the manifold of
substructural shapes in ${{\mathbb{R}}^{k}}$. We denote by $\mathcal{V}_{\nu
}$ the subspace of $PA\left( \mathcal{T}\right) \otimes H^{1}\left( I,{{%
\mathbb{R}}^{k}}\right) $ which consists of all mappings $\nu $ satisfying $%
\nu \left\vert _{\partial \mathcal{B}_{\nu }}\right. =0$. Furthermore, we
set $M=k\#{\mathcal{T}}$.

The spatial discretization $\mathcal{A}_{\mathcal{T}}$ of $\mathcal{A}$ is
then obtained by restricting to $\mathcal{V}_{\mathcal{T}}:=\mathcal{V}%
_{u}\times \mathcal{V}_{\nu }$ the domain of $\mathcal{A}$. More precisely,
we consider the semi-discrete action $\mathcal{A}_{\mathcal{T}}:PA\left( 
\mathcal{T}\right) \otimes H^{1}\left( I,{\mathbb{R}}^{d\times k}\right)
\rightarrow \lbrack 0,+\infty ]$ defined by 
\begin{equation*}
\mathcal{A}_{\mathcal{T}}(u,\nu ):=%
\begin{cases}
\mathcal{A}(u,\nu ) & \text{ if }(u,\nu )\in \mathcal{V}_{\mathcal{T}}\cr%
+\infty & \text{ otherwise. }%
\end{cases}%
\end{equation*}%
Notice that the nodal pair $(u(\cdot ),\nu (\cdot ))$ belongs to $%
H^{1}\left( I,{{\mathbb{R}}^{D\times M}}\right) $ for every map $(u,\nu )\in 
\mathcal{V}_{\mathcal{T}}$. Hence, the expansion in (\ref{Shape}) allows one
to regard $\mathcal{A}_{\mathcal{T}}$ as a functional defined over $%
H^{1}\left(I,{{\mathbb{R}}^{D\times M}}\right) $. We agree to do that in the
sequel in order to simplify the notation.

A further simplification concerns the referential mass density $\rho $. In
the sequel we simply presume that $\rho $ is constant along the motion.
Moreover, we assume that the mass matrix associated with the spatial
discretization can be expressed in diagonal form in some basis. Then, a
straightforward computation in that basis leads to 
\begin{equation*}
\mathcal{A}_{\mathcal{T}}\left( \left[ t_{0},t_{f}\right] ;u,\nu \right)
:=\sum_{K\in \mathcal{T}}\int_{I}\mathcal{A}\left( K;(u_{K},{\nu _{K}}%
)(t)\right) \,dt,
\end{equation*}%
where, for each $K\in \mathcal{T}$ , 
\begin{equation*}
\mathcal{A}\left( (u_{K},{\nu _{K}})(t),K\right) :=\sum_{a\in K}\left( \frac{%
m_{K,a}}{2}\left\vert \dot{u}_{a}(t)\right\vert ^{2}+\frac{\overline{\rho }%
_{K,a}}{2}\left\vert \dot{\nu}_{a}(t)\right\vert ^{2}\right) -V_{K}((u_{K},{%
\nu _{K}})(t)),
\end{equation*}%
and 
\begin{eqnarray}
\lefteqn{V_{K}\left( (u_{K},\nu _{K})(t)\right) :=\int_{K}e\left( (\nabla
u_{K},\nu _{K},\nabla \nu _{K})(x,t)\right) \,dx}  \label{VuKappa} \\
&&+\int_{K}w\left( (u_{K},{\nu _{K}})(x,t)\right) \,dx-\int_{\partial K\cap
\partial \mathcal{B}_{t}}\mathsf{t}(x)\cdot u_{K}(x,t)\,d\mathcal{H}^{d-1}. 
\notag
\end{eqnarray}%
The diagonalization of the mass matrix leaves invariant the Lagrangian: it
can be considered, in a certain sense, as a change in observer.

We call \emph{stationary point} for $\mathcal{A}_{\mathcal{T}}$ any element $%
(u,\nu )$ of $H^{1}\left( I,{{\mathbb{R}}^{D\times M}}\right) $ which for
any $(\varphi ,\phi )\in H_{0}^{1}\left( I,{{\mathbb{R}}^{D\times M}}\right) 
$, with $\varphi _{a}\equiv 0$ if $a\in \partial \mathcal{B}_{u}$ and $\phi
_{a}\equiv 0$ if $a\in \partial \mathcal{B}_{\nu }$, satisfies the
d'Alembert-Lagrange-type principle 
\begin{equation}
\delta \mathcal{A}_{\mathcal{T}}(u,\nu )[\varphi ,\phi ]+\mathcal{D}_{%
\mathcal{T}}({\dot{\nu}})[\phi ]=0.  \label{ddAL}
\end{equation}%
The semi-discretized dissipation $\mathcal{D}_{\mathcal{T}}$ that appears
above is defined by 
\begin{equation*}
\mathcal{D}_{\mathcal{T}}({\dot{\nu}})[\phi ]:=\eta \sum_{K\in \mathcal{T}%
}\int_{I}{\dot{\nu}}_{K}\cdot \phi _{K}\,dt,
\end{equation*}%
where $\phi _{K}$ is the restriction of $\phi $ to $K$.

The d'Alembert-Lagrange-type principle \eqref{ddAL} is equivalent to the
requirement that for any node $a\in \mathcal{T}\setminus (\partial \mathcal{B%
}_{u}\cup \partial \mathcal{B}_{\nu })$, and any $(\varphi _{a},\phi
_{a})\in (u_{a},\nu _{a})+H_{0}^{1}\left( I,{\mathbb{R}}^{d\times k}\right) $%
, if $m_{a}:=\sum_{\left\{ K|a\in K\right\} }m_{K,a}$ is the \emph{nodal mass%
} of node $a$ and $\overline{\rho }_{a}:=\sum_{\left\{ K|a\in K\right\} }%
\overline{\rho }_{K,a}$, we have 
\begin{eqnarray}
\lefteqn{\int_{I}m_{a}\dot{u}_{a}\left( t\right) \cdot \dot{\varphi}%
_{a}(t)\,dt}  \notag  \label{dALL1} \\
&=&\int_{I}\sum_{\left\{ K|a\in K\right\} }\left( \int_{K}\partial
_{F}e\left( (\nabla u_{K},{\nu _{K}},\nabla {\nu _{K}})(x,t)\right) \nabla 
\mathcal{N}_{a}(x)\,dx\right) \cdot \varphi _{a}(t)\,dt  \notag \\
&&+\int_{I}\sum_{\left\{ K|a\in K\right\} }\left( \int_{K}\partial
_{u}w\left( (u_{K},{\nu _{K}})(x,t)\right) \mathcal{N}_{a}(x)dx\right) \cdot
\varphi _{a}(t)\,dt  \notag \\
&&-\int_{I}\sum_{\left\{ K|a\in K\right\} }\left( \int_{\partial K\cap
\partial \mathcal{B}_{t}}\mathsf{t}(x)\mathcal{N}_{a}(x)\,d\mathcal{H}%
^{d-1}(x)\right) \cdot \varphi _{a}(t)\,dt,
\end{eqnarray}%
and 
\begin{eqnarray}
\lefteqn{\int_{I}\overline{\rho }_{a}\dot{\nu}_{a}(t)\cdot \dot{\phi}%
_{a}(t)\,dt+\int_{I}\eta _{a}\dot{\nu}_{a}(t)\cdot \phi _{a}(t)dt\notag}
\label{dALL2} \\
&=&\int_{I}\sum_{\{K|a\in K\}}\left( \int_{K}\partial _{N}e\left( (\nabla
u_{K},{\nu _{K}},\nabla {\nu _{K}})(x,t)\right) \nabla \mathcal{N}%
_{a}(x)\,dx\right) \cdot \phi _{a}(t)\,dt  \notag \\
&&+\int_{I}\sum_{\{K|a\in K\}}\left( \int_{K}\partial _{\nu }e\left( (\nabla
u_{K},{\nu _{K}},\nabla {\nu _{K}})(x,t)\right) \mathcal{N}%
_{a}(x)\,dx\right) \cdot \phi _{a}(t)\,dt  \notag \\
&&+\int_{I}\sum_{\{K|a\in K\}}\left( \int_{K}\partial _{\nu }w\left( (u_{K},{%
\nu _{K}})(x,t)\right) \mathcal{N}_{a}(x)\,dx\right) \cdot \phi _{a}(t)\,dt.
\end{eqnarray}%
Moreover, in case $a\in \partial \mathcal{B}_{u}\backslash \partial \mathcal{%
B}_{\nu }$ only equation \eqref{dALL2} has to be satisfied since $%
u_{a}(t)\equiv 0$, while if $a\in \partial \mathcal{B}_{\nu }\backslash
\partial \mathcal{B}_{u}$ we have $\nu _{a}(t)\equiv 0$ and the sole %
\eqref{dALL1} has to be satisified.

Results about finite elements for complex bodies imply the theorem below
(see examples in \cite{MS}, \cite{GMS}).

\begin{theorem}
Assume (A1)-(A3) and consider a family $(\mathcal{T}_m)$ of regular
triangulations of $\mathcal{B}$, with $m>0$ the mesh size of $\mathcal{T}_m$%
. Let also $(u_m,\nu_m)\in \mathcal{V}_{\mathcal{T}_{m}}$ be a stationary
point for $\mathcal{A}_{\mathcal{T}_{m}} $. The sequence $((u_m,\nu_m))_{m\in%
\mathbb{N}}$ converges in $H^{1}\left(\mathcal{B}\times I,{{\mathbb{R}}%
^{d\times k}}\right)$ to a stationary point of $\mathcal{A}$.
\end{theorem}

\subsection{Time discretization}

For the discretization of the time interval $I=(t_{0},t_{f})$ a partition $%
\Theta :=\left\{ t_{i}\right\} _{i=0,...,N_{\Theta }}$ of $[t_{0},t_{f}]$
with $t_{N_{\Theta }}=t_{f}$ is selected. Its \emph{size} $h_{\Theta }$ is
the number 
\begin{equation*}
h_{\Theta }:=\max_{i}\left( t_{i+1}-t_{i}\right) .
\end{equation*}
Each $K\in \mathcal{T}$ is endowed with an \emph{elemental time set }which
is an ordered subset $\Theta _{K}$ of $\Theta $. By relabeling the elements
we write%
\begin{equation*}
\Theta _{K}=\left\{
t_{0}=t_{K}^{0}<...<t_{K}^{N_{K}-1}<t_{K}^{N_{K}}=t_{f}\right\} .
\end{equation*}%
We assume $\Theta =\cup _{K\in \mathcal{T}}\Theta _{K}$ and, for the sake of
simplicity, that $\Theta _{K}\cap \Theta _{K^{\prime }}=\{t_{0},t_{f}\}$ for
any $K$, $K^{\prime }\in \mathcal{T}$ with $K\neq K^{\prime }$. We denote by 
$T_{\Theta }$ the maximum of the elemental time sizes, namely 
\begin{equation*}  \label{eltimesize}
T_{\Theta }:=\max_{K}\max_{\Theta _{K}}(t_{K}^{j+1}-t_{K}^{j}).
\end{equation*}
The circumstance that each finite element can be endowed with a different
time set is the basic characteristic of AVIs as already mentioned: if one
imposes appropriate choices of elemental time sets, one may prove
conservation of energy in discrete time. Take note that in principle such
choices could not exist (see related comments in \cite{LMOW1}).

For a node $a$ in $\mathcal{T}$, elemental time sets define also the \emph{%
nodal time set} $\Theta _{a}$ by 
\begin{equation*}
\Theta _{a}:=\bigcup_{\left\{ K\,|\,a\in K\right\} }\Theta _{K}=\left\{
t_{0}=t_{a}^{1}<...<t_{a}^{N_{a}-1}<t_{a}^{N_{a}}=t_{f}\right\} .
\end{equation*}%
As a measure of the \emph{asynchronicity of} $\Theta $, we consider the
ratio 
\begin{equation}
\tau _{\Theta }:=\frac{\max_{K\in \mathcal{T}}\left( \max_{\Theta
_{K}}\left( t_{K}^{j+1}-t_{K}^{j}\right) \right) }{\min_{K\in \mathcal{T}%
}\left( \min_{\Theta _{K}}\left( t_{K}^{j+1}-t_{K}^{j}\right) \right) },
\label{MTeta}
\end{equation}%
and notice that \footnote{%
Indeed, by definition of time size we have $h_{\Theta }\leq T_{\Theta }$ and 
$t_{f}-t_{0}\leq h_{\Theta }\#\Theta $. For every element $K$ we have $%
\min_{K\in {\mathcal{T}}}(\min_{\Theta _{K}}(t_{K}^{j+1}-t_{K}^{j}))\#\Theta
_{K}\leq %
t_{f}-t_{0}$, then by summing on $K$ it follows that 
\begin{equation*}
\min_{K\in {\mathcal{T}}}(\min_{\Theta _{K}}(t_{K}^{j+1}-t_{K}^{j}))\leq
h_{\Theta }\#\mathcal{T}.
\end{equation*}%
Inequality (\ref{nuova}) then follows by definition of $\tau _{\Theta }$
(see (\ref{MTeta})).} 
\begin{equation}
h_{\Theta }\leq T_{\Theta }\leq \tau _{\Theta }h_{\Theta }\#\mathcal{T}.
\label{nuova}
\end{equation}

In the sequel we assume that each node $a\in \mathcal{T}$ follows a \emph{%
linear} trajectory within each time interval with end-points that are
consecutive instants in $\Theta_{a}$. Such a choice characterizes the class
of AVIs that we analyze here. Then we denote by $\mathcal{Y}_{\Theta}$ the
subspace of functions in $H^{1}\left(I,{{\mathbb{R}}^{D\times M}}\right)$
which are continuous and with piecewise constant time rates in the intervals
in $\Theta_{a}$, and such that $u_a\equiv 0$ if $a\in\partial\mathcal{B}_u$
and $\nu_a\equiv 0$ if $a\in\partial\mathcal{B}_\nu$. Thus, for each $%
(u,\nu)\in\mathcal{Y}_{\Theta}$, $a\in \mathcal{T}$, $t_{a}^{i}\in \Theta
_{a}$ and $t\in \left[ t_{a}^{i},t_{a}^{i+1}\right) $ we set 
\begin{equation*}
(\dot{u}_{a},\dot{\nu}_{a})(t)=\left(\frac{u_{a}\left( t_{a}^{i+1}\right)
-u_{a}\left( t_{a}^{i}\right) }{t_{a}^{i+1}-t_{a}^{i}}, \frac{\nu_{a}\left(
t_{a}^{i+1}\right) -\nu_{a}\left( t_{a}^{i}\right) }{t_{a}^{i+1}-t_{a}^{i}}%
\right).
\end{equation*}
By following \cite{LMOW1}, the fully discrete action sum in time is defined
for $(u,\nu)\in\mathcal{Y}_{\Theta}$ by 
\begin{equation}  \label{discractsum}
\mathcal{A}_{\mathcal{T},\Theta }\left(u,\nu\right):= \sum_{K\in \mathcal{T}%
}\sum_{j=0}^{N_K-1} \mathcal{A}^{j}\left(K;u_{K},{\nu_K}\right),
\end{equation}
where for the generic finite element $K$, $\mathcal{A}^{j}\left(K;u_{K},{%
\nu_K}\right) $ is defined by 
\begin{eqnarray*}
\mathcal{A}^{j}&&\hskip-0.7cm\left(K;u_{K},{\nu_K}\right):=\sum_{a\in K}
\sum_{\left\{ i|t_{a}^{i}\in \left[ t_{K}^{j},t_{K}^{j+1}\right) \right\}}%
\frac {m_{K,a}}2\left( t_{a}^{i+1}-t_{a}^{i}\right) \left\vert\dot{u}%
_{a}\left( t_{a}^{i}\right) \right\vert ^{2} \\
&&\hskip-0.7cm+\sum_{a\in K} \sum_{\left\{ i|t_{a}^{i}\in \left[
t_{K}^{j},t_{K}^{j+1}\right) \right\}}\frac {\overline{\rho}_{K,a}}2\left(
t_{a}^{i+1}-t_{a}^{i}\right) \left\vert\dot{\nu}_{a}\left(t_{a}^{i}\right)%
\right\vert^{2} -(t_{K}^{j+1}-t_{K}^{j})V_{K}((u_{K},{\nu_K})(t_{K}^{j})),
\end{eqnarray*}
with $V_{K}$ given by (\ref{VuKappa}). The choice of the approximation 
\begin{eqnarray*}
\int_{t_{K}^{j}}^{t_{K}^{j+1}}V_K\left((u_K,{\nu_K})(t)\right)\,dt\approx
(t_{K}^{j+1}-t_{K}^{j})V_{K}((u_{K},{\nu_K})(t_{K}^{j}))
\end{eqnarray*}
gives rise to explicit integrators of central-difference type and is only
one of the possible schemes that can be used.

It is also convenient to define all action sums on the same function space
to avoid to link the function space itself to the choice of $\Theta $. For
this reason we extend $\mathcal{A}_{\mathcal{T},\Theta }$ to the whole $%
\mathcal{Y}:=H^{1}\left(I,{{\mathbb{R}}^{D\times M}}\right)$. With a little
abuse of notation we set 
\begin{eqnarray*}
\mathcal{A}_{\mathcal{T},\Theta}\left(u,\nu\right)=%
\begin{cases}
\text{ as in } \eqref{discractsum} & \text{ if } (u,\nu)\in \mathcal{Y}%
_{\Theta}\cr +\infty & \text{ if } (u,\nu)\in \mathcal{Y}\backslash \mathcal{%
Y}_{\Theta}.%
\end{cases}%
\end{eqnarray*}
The \emph{discrete stationary points} of $\mathcal{A}_{\mathcal{T},\Theta}$
are couples $(u,\nu)\in \mathcal{Y}_{\Theta}$ satisfying the fully discrete
analogous of the variational principle \eqref{ddAL}. More precisely, for all
test functions $(\varphi,\phi)\in \mathcal{Y}_{\Theta}\cap H^{1}_0(I,{{%
\mathbb{R}}^{D\times M}})$ there holds 
\begin{equation}  \label{dALdiscr}
\delta \mathcal{A}_{\mathcal{T},\Theta}(u,\nu)[\varphi,\phi]+ \mathcal{D}_{%
\mathcal{T},\Theta}({\dot{\nu}})[\phi]=0,
\end{equation}
where the \emph{fully discretized dissipation} is given by 
\begin{equation}  \label{dissdiscr}
\mathcal{D}_{\mathcal{T},\Theta}({\dot{\nu}})[\phi]:= \sum_{a\in\mathcal{T}%
}\sum_{i=1}^{N_a}\eta_a
(\nu_a(t_a^i)-\nu_a(t_a^{i-1}))\cdot\phi_a(t_a^{i-1}).
\end{equation}
Alternatively, we will refer to the discrete stationary points also as to 
\emph{discrete solutions}.

For $\nu$ and $\phi$ as above, by a direct computation we infer 
\begin{equation}  \label{dissdiscr1}
\mathcal{D}_{\mathcal{T}}({\dot{\nu}})[\phi] =\sum_{a\in\mathcal{T}%
}\sum_{i=1}^{N_a}\eta_a (\nu_a(t_a^i)-\nu_a(t_a^{i-1}))\cdot\frac{%
\phi_a(t_a^i)+\phi(t_a^{i-1})}2.
\end{equation}
The discretized form of the dissipation has been chosen to be %
\eqref{dissdiscr} for the sake of simplicity. Furthermore, notice that by %
\eqref{dissdiscr} and \eqref{dissdiscr1} we have 
\begin{eqnarray}  \label{diffdiss}
\lefteqn{\left|\mathcal{D}_{\mathcal{T}}(\dot{\nu})[\phi]- \mathcal{D}_{%
\mathcal{T},\Theta}(\dot{\nu})[\phi]\right|= \left| \sum_{a\in\mathcal{T}%
}\sum_{i=1}^{N_a}\eta_a(\nu_a(t_a^i)-\nu_a(t_a^{i-1})) \cdot\frac{%
\phi_a(t_a^i)-\phi(t_a^{i-1})}2\right|}  \notag \\
&& =\frac 12\left|\sum_{a\in\mathcal{T}%
}\sum_{i=1}^{N_a}(t_a^i-t_a^{i-1})^{2} \eta_a\dot{\nu}_a(t_a^{i-1})\cdot\dot{%
\phi}(t_a^{i-1})\right| \leq\frac {\eta T_\Theta}2 \left|\int_{\mathcal{B}%
\times I}\dot{\nu}\cdot\dot{\phi}\,dtdx\right|.
\end{eqnarray}

In Proposition~\ref{existence} below we will show that given initial
conditions $(u(t_{0}),\nu (t_{0}))$ and $(\dot{u}(t_{0}),{\dot{\nu}}(t_{0}))$%
, the discrete d'Alembert-Lagrange-type variational principle %
\eqref{dALdiscr} defines inductively a unique trajectory $(u,\nu )\in 
\mathcal{Y}_{\Theta }$.

\subsection{Convergence in Time}

The following theorem clarifies the stage.

\begin{theorem}
\label{main} Assume (A1)-(A3). Let $(\Theta _{h})_{h\in {\mathbb{N}}}$ be a
sequence of time sets for a bounded time interval $I=(t_{0},t_{f})$ such
that $T_{\Theta _{h}}\rightarrow 0$ as $h\rightarrow +\infty $. Let also $%
(u_{h}(t_{0}),\nu _{h}(t_{0}))$ and $(\dot{u}_{h}(t_{0}),\dot{\nu}%
_{h}(t_{0}))$ be initial conditions satisfying 
\begin{equation}
\sup_{h}\left( \tau _{\Theta _{h}}+\left\vert u_{h}(t_{0})\right\vert
+\left\vert \nu _{h}(t_{0})\right\vert +\left\vert \dot{u}%
_{h}(t_{0})\right\vert +\left\vert \dot{\nu}_{h}(t_{0})\right\vert \right)
<+\infty .  \label{incondbound}
\end{equation}%
Then the discrete d'Alembert-Lagrange-type principle \eqref{dALdiscr}
relative to $\Theta _{h}$ has a unique solution $(u_{h},\nu _{h})$.

Moreover, there exists $(u,\nu )\in W^{2,\infty} (I,{{\mathbb{R}}^{D\times M}%
})$ satisfying \eqref{ddAL} such that, up to a sub-sequence, $%
((u_{h},\nu_{h}))_{h\in \mathbb{N}}$ converges to $(u,\nu )$ uniformly on $%
\overline{I}$, and $((\dot{u}_{h},\dot{\nu}_{h}))_{h\in {\mathbb{N}}}$
converges to $(\dot{u},\dot{\nu})$ in $L^{p}(I,{{\mathbb{R}}^{D\times M}})$
for every $p\in\lbrack 1,+\infty )$.
\end{theorem}

To prove the theorem above, preliminary results are necessary. As a first
step by Lemma~\ref{ddALexpl} we establish an explicit form of the discrete
variational principle~(\ref{dALdiscr}). From such a form, it is easy to
infer existence and uniqueness of discrete solutions with given initial
conditions. Then we show $BV$ estimates for the velocities of discrete
solutions in Proposition~\ref{stime}.

By making variations that fix all the nodes except one at a given time, one
can find easily the following equivalent form of the
d'Alembert-Lagrange-type principle \eqref{dALdiscr}.

\begin{lemma}
\label{ddALexpl} Let $(u,\nu )\in \mathcal{Y}_{\Theta }$, then $(u,\nu )$
solves \eqref{dALdiscr} if and only if given any node $a\in {\mathcal{T}}$
and any nodal time $t_{a}^{i}\in \left( t_{0},t_{f}\right] $, denoting by $K$
the sole element for which $t_{a}^{i}\in \Theta _{K}$ and $%
t_{K}^{j}=t_{a}^{i}$, if $a\in \mathcal{T}\backslash \left( \partial 
\mathcal{B}_{u}\cup \partial \mathcal{B}_{\nu }\right) $, the following
balance equations hold: 
\begin{eqnarray}
\lefteqn{m_{a}\left( \dot{u}_{a}\left( t_{a}^{i-1}\right) -\dot{u}_{a}\left(
t_{a}^{i}\right) \right) \notag}  \label{Discr-dAL1} \\
&=&(t_{K}^{j+1}-t_{K}^{j})\int_{K}\partial _{F}e\left( (\nabla u_{K},{\nu
_{K}},\nabla {\nu _{K}})(x,t_{K}^{j})\right) \nabla \mathcal{N}_{a}(x)\,dx 
\notag \\
&&+(t_{K}^{j+1}-t_{K}^{j})\int_{K}\partial _{u}w\left( (u_{K},{\nu _{K}}%
)(x,t_{K}^{j})\right) \mathcal{N}_{a}(x)\,dx  \notag \\
&&-(t_{K}^{j+1}-t_{K}^{j})\int_{\partial K\cap \partial \mathcal{B}_{t}}%
\mathsf{t}(x)\mathcal{N}_{a}(x)\,d\mathcal{H}^{d-1},
\end{eqnarray}%
and 
\begin{eqnarray}
\lefteqn{\overline{\rho }_{a}\left( \dot{\nu}_{a}\left( t_{a}^{i-1}\right) -%
\dot{\nu}_{a}\left( t_{a}^{i}\right) \right) +\eta _{a}(\nu
_{a}(t_{a}^{i+1})-\nu _{a}(t_{a}^{i}))}  \notag  \label{Discr-dAL2} \\
&=&(t_{K}^{j+1}-t_{K}^{j})\int_{K}\partial _{N}e\left( (\nabla u_{K},{\nu
_{K}},\nabla {\nu _{K}})(x,t_{K}^{j})\right) \nabla \mathcal{N}_{a}(x)\,dx 
\notag \\
&&+(t_{K}^{j+1}-t_{K}^{j})\int_{K}\partial _{\nu }e\left( (\nabla u_{K},{\nu
_{K}},\nabla {\nu _{K}})(x,t_{K}^{j})\right) \mathcal{N}_{a}(x)\,dx  \notag
\\
&&+(t_{K}^{j+1}-t_{K}^{j})\int_{K}\partial _{\nu }w\left( (u_{K},{\nu _{K}}%
)(x,t_{K}^{j})\right) \mathcal{N}_{a}(x)\,dx.
\end{eqnarray}%
Moreover, if $a\in \partial \mathcal{B}_{u}\backslash \partial \mathcal{B}%
_{\nu }$, only the first equation has to be satisfied, while, if $a\in
\partial \mathcal{B}_{\nu }\backslash \partial \mathcal{B}_{u}$, only the
second equation has to be satisfied.
\end{lemma}

Clearly, equations \eqref{Discr-dAL1} and \eqref{Discr-dAL2} define
inductively a unique discrete trajectory in $\mathcal{Y}_\Theta$ once
initial conditions have been specified.

Estimates on the $L^{\infty }$ norm on the time interval $I$ of the velocity
of stationary points can be derived by exploiting directly the discrete
stationarity conditions and the growth conditions of the potential energy
densities as suggested in \cite{MM} (see also \cite{FM}). Here, we provide
also an estimate on the pointwise variation of the velocities of stationary
points which has been overlooked in previous works. In turn this estimate
let us gain strong compactness properties and permits us to pass to the
limit directly into equations \eqref{Discr-dAL1} and \eqref{Discr-dAL2}, or
equivalently in \eqref{dALdiscr}. This method is alternative to the ones
already developed to study the convergence properties of variational
integrators (see \cite{LMOW2}, \cite{MO}, \cite{MM}, \cite{FM}).
Furthermore, the $BV$ estimate wil be instrumental in Section~\ref{VIs} (see
the related comments there).

\begin{proposition}
\label{stime} There exist a constant $T_0=T_0(\eta,\overline{\rho}%
,\lambda,\Lambda,\mathcal{T})>0$ such that given initial conditions $%
\left(u(t_0),\nu(t_0)\right)$ and $\left(\dot{u}(t_0),\dot{\nu}(t_0)\right)$
for every entire time set $\Theta $ with $T_\Theta\in(0,T_0)$, the solution $%
(u,\nu)\in \mathcal{Y}_{\Theta }$ of equations \eqref{Discr-dAL1} and %
\eqref{Discr-dAL2} satisfies 
\begin{equation}  \label{Prop1}
\|\dot{u}\| _{L^{\infty }\left(I,{{\mathbb{R}}^D}\right)}+ \|\dot{\nu}\|
_{L^{\infty }\left(I,{{\mathbb{R}}^M}\right)} \leq \kappa\exp \left(
\kappa\tau_\Theta\right),
\end{equation}
for some constant $\kappa>0$ depending on the initial conditions themselves
and on the data of the problem.

Moreover, the functions $\dot{u}$, ${\dot{\nu}}$ have (pointwise) bounded
variation $pV$ on $\overline{I}$, with 
\begin{equation}  \label{stimaBV}
pV(\dot{u},[t_1,t_2])+pV(\dot{\nu},[t_1,t_2]) \leq
\kappa\tau_\Theta\left(1+\exp\left(\kappa\tau_\Theta\right)\right)
(t_2-t_1+2T_\Theta),
\end{equation}
for every interval $[t_1,t_2]\subseteq \overline{I}$.
\end{proposition}

\begin{proof}
Fix a node $a$ in the tessellation $\mathcal{T}$ and a nodal time $%
t_{a}^{i}\in \Theta _{a}$. By assumption there exists a unique $K\in 
\mathcal{T}$ such that $t_{a}^{i}\in \Theta _{K}$, with $t_{a}^{i}=t_{K}^{j}$%
. The definition of nodal time set yields $t_{a}^{i}=t_{\alpha }^{i(\alpha
)}\in \Theta _{\alpha }$ for all the other nodes $\alpha \in K$. Moreover,
recalling that $\Theta =\{t_{l}\}_{l=1,\ldots ,N_{\Theta }}$, we assume $%
t_{a}^{i}=t_{s+1}\in \Theta $ for some $s$.

We first estimate the velocity of $u$ at time $t_{s+1}$ in terms of the
velocities of $u$ itself and $\nu $ at previous times. In doing that we use
the discrete d'Alembert-Lagrange-type equation \eqref{Discr-dAL1},
assumptions (A1)-(A3), and \eqref{stimagrad} which entail 
\begin{equation}
m_{a}|\dot{u}_{a}\left( t_{a}^{i}\right) -\dot{u}_{a}\left(
t_{a}^{i-1}\right) |\leq c(t_{K}^{j+1}-t_{K}^{j})\left( 1+\sum_{\alpha \in
K}(|u_{\alpha }(t_{\alpha }^{i(\alpha )})|+|\nu _{\alpha }(t_{\alpha
}^{i(\alpha )})|)\right) .  \label{stimaBV1}
\end{equation}%
By taking into account that the maps $u_{\alpha },\nu _{\alpha }$ are
piecewise affine in time, we get 
\begin{eqnarray*}
\lefteqn{\left\vert \dot{u}_{a}\left( t_{a}^{i}\right) -\dot{u}_{a}\left(
t_{a}^{i-1}\right) \right\vert \leq c(t_{K}^{j+1}-t_{K}^{j})\left(
1+\sum_{\alpha \in K}\left( \left\vert u_{\alpha }(t_{0})\right\vert
+\left\vert \nu _{\alpha }(t_{0})\right\vert \right) \right) } \\
&&+c(t_{K}^{j+1}-t_{K}^{j})\sum_{\alpha \in K}\sum_{l=0}^{i(\alpha
)-1}\left( t_{\alpha }^{l+1}-t_{\alpha }^{l}\right) (\left\vert \dot{u}%
_{\alpha }\left( t_{\alpha }^{l}\right) \right\vert +\left\vert \dot{\nu}%
_{\alpha }\left( t_{\alpha }^{l}\right) \right\vert ) \\
&\leq &cT_{\Theta }+cT_{\Theta }(t_{f}-t_{0})\sum_{\alpha \in K}\left( \Vert 
\dot{u}_{\alpha }\Vert _{L^{\infty }\left( (t_{0},t_{\alpha }^{i(\alpha
)-1}),{{\mathbb{R}}^{d}}\right) }+\Vert \dot{\nu}_{\alpha }\Vert _{L^{\infty
}\left( (t_{0},t_{\alpha }^{i\left( \alpha \right) -1}),{{\mathbb{R}}^{k}}%
\right) }\right) . \\
&&
\end{eqnarray*}%
By recalling that $t_{a}^{i}=t_{s+1}$, and since $t_{\alpha }^{i\left(
\alpha \right) -1}\leq t_{s}$ for every $\alpha \in K$, the latter
inequality yields 
\begin{equation}
\left\Vert \dot{u}\right\Vert _{L^{\infty }\left( (t_{0},t_{s+1}),{{\mathbb{R%
}}^{D}}\right) }\leq cT_{\Theta }\left( 1+\left\Vert \dot{\nu}\right\Vert
_{L^{\infty }\left( (t_{0},t_{s}),{{\mathbb{R}}^{M}}\right) }\right)
+(1+cT_{\Theta })\left\Vert \dot{u}\right\Vert _{L^{\infty }\left(
(t_{0},t_{s}),{{\mathbb{R}}^{D}}\right) }.  \label{stimaLinf1}
\end{equation}%
We prove the analogous estimate for ${\dot{\nu}}$, the only difference with
the previous argument being that we have to take into account, in addition
to the previous case, for the presence of the discretized dissipation. Thus
from \eqref{Discr-dAL2}, assumptions (A1)-(A3) and \eqref{stimagrad} we get 
\begin{equation}
\overline{\rho }_{a}|\dot{\nu}_{a}\left( t_{a}^{i}\right) -\dot{\nu}%
_{a}\left( t_{a}^{i-1}\right) |\leq \eta _{a}T_{\Theta }|\dot{\nu}_{a}\left(
t_{a}^{i}\right) |+cT_{\Theta }\left( 1+\sum_{\alpha \in K}(|u_{\alpha
}(t_{\alpha }^{i(\alpha )})|+|\nu _{\alpha }(t_{\alpha }^{i(\alpha
)})|)\right) ,  \label{stimaBV1bis}
\end{equation}%
which in turn yields 
\begin{eqnarray*}
\lefteqn{\left\vert \dot{\nu}_{a}\left( t_{a}^{i}\right) -\dot{\nu}%
_{a}\left( t_{a}^{i-1}\right) \right\vert \leq cT_{\Theta }|\dot{\nu}%
_{a}\left( t_{a}^{i}\right) |+cT_{\Theta }} \\
&&+cT_{\Theta }(t_{f}-t_{0})\sum_{\alpha \in K}\left( \Vert \dot{u}_{\alpha
}\Vert _{L^{\infty }\left( (t_{0},t_{\alpha }^{i(\alpha )-1}),{{\mathbb{R}}%
^{d}}\right) }+\Vert \dot{\nu}_{\alpha }\Vert _{L^{\infty }\left(
(t_{0},t_{\alpha }^{i\left( \alpha \right) -1}),{{\mathbb{R}}^{k}}\right)
}\right) .
\end{eqnarray*}%
Thus, there exists a constant $T_{0}=T_{0}(\eta ,\overline{\rho}%
,\lambda,\Lambda,\mathcal{T})>0$ such that for every $T_{\Theta }\in
(0,T_{0})$ we have 
$(1-cT_{\Theta })\geq 1/2$. Finally, we deduce 
\begin{equation}
\left\Vert \dot{\nu}\right\Vert _{L^{\infty }\left( (t_{0},t_{s+1}),{{%
\mathbb{R}}^{M}}\right) }\leq cT_{\Theta }\left( 1+\left\Vert \dot{u}%
\right\Vert _{L^{\infty }\left( (t_{0},t_{s}),{{\mathbb{R}}^{D}}\right)
}\right) +(1+cT_{\Theta })\left\Vert \dot{\nu}\right\Vert _{L^{\infty
}\left( (t_{0},t_{s}),{{\mathbb{R}}^{M}}\right) }.  \label{stimaLinf2}
\end{equation}%
In particular, by setting $\beta _{l}:=\Vert \dot{u}\Vert _{L^{\infty
}\left( (t_{0},t_{l}),{{\mathbb{R}}^{D}}\right) }+\Vert \dot{\nu}\Vert
_{L^{\infty }\left( (t_{0},t_{l}),{{\mathbb{R}}^{M}}\right) }$, from (\ref%
{stimaLinf1}), (\ref{stimaLinf2}) we infer 
\begin{equation*}
\beta _{s+1}\leq cT_{\Theta }+\left( 1+cT_{\Theta }\right) \beta _{s}.
\end{equation*}%
Such inequality yields by iteration 
\begin{eqnarray}
\lefteqn{\beta _{s+1}\leq cT_{\Theta }\sum_{i=0}^{s}\left( 1+cT_{\Theta
}\right) ^{i}+\beta _{0}\left( 1+cT_{\Theta }\right) ^{s+1}}
\label{stimaglob} \\
&\leq &\left( 1+\beta _{0}\right) \left( 1+cT_{\Theta }\right) ^{s+1}\leq
c\left( 1+cT_{\Theta }\right) ^{\#\Theta }.  \notag
\end{eqnarray}%
%
%
%
%
Eventually, since $T_{\Theta }\#\Theta \leq (t_{f}-t_{0})\tau _{\Theta }\#%
\mathcal{T}$, with $\tau _{\Theta }$ defined in (\ref{MTeta}), from
inequality \eqref{stimaglob} we deduce 
\begin{equation*}
\Vert \dot{u}\Vert _{L^{\infty }\left( I,{{\mathbb{R}}^{D}}\right) }+\Vert 
\dot{\nu}\Vert _{L^{\infty }\left( I,{{\mathbb{R}}^{M}}\right) }=\beta
_{\#\Theta }\leq c\left( 1+cT_{\Theta }\right) ^{c\tau _{\Theta }/T_{\Theta
}},
\end{equation*}%
and then (\ref{Prop1}) follows.

In order to prove the $BV$ estimate \eqref{stimaBV} we note first that since 
$\dot{u}_a$, $\dot{\nu}_a$ are piecewise constant, their (pointwise)
variation over an interval $[t_1,t_2]\subseteq \overline{I}$ is given by the
sum of the jumps in $[t_1,t_2]$. Hence, by using \eqref{stimaBV1} and %
\eqref{stimaBV1bis}, and taking advantage of \eqref{Prop1}, we obtain 
\begin{equation}  \label{stimaBV3}
\left\vert\dot{u}_{a}\left(t_{a}^{i}\right)- \dot{u}_{a}\left(t_{a}^{i-1}%
\right)\right\vert +\left\vert\dot{\nu}_{a}\left(t_{a}^{i}\right)- \dot{\nu}%
_{a}\left(t_{a}^{i-1}\right)\right\vert \leq cT_\Theta
(1+c\exp(c\tau_\Theta)).
\end{equation}
To compute the pointwise variation of $\dot{u}_a,\,{\dot{\nu}}_a$ on $%
[t_1,t_2]$ we sum \eqref{stimaBV3} over the set of indices $%
\Sigma_{t_1,t_2}:=\{i|\,[t_{a}^{i-1},t_{a}^i]\cap[t_1,t_2]\neq\emptyset\}$.
Furthermore, by taking into account that $T_\Theta\#\Sigma_{t_1,t_2}\leq
c\tau_\Theta(t_2-t_1+2T_\Theta)$, we conclude that 
\begin{eqnarray*}
\lefteqn{pV(\dot{u}_a,[t_1,t_2])+pV(\dot{\nu}_a,[t_1,t_2])} \\
&&= \sum_{i\in\Sigma_{t_1,t_2}} \left(\left\vert\dot{u}_{a}\left(t_{a}^{i}%
\right)- \dot{u}_{a}\left(t_{a}^{i-1}\right)\right\vert+ \left\vert\dot{\nu}%
_{a}\left(t_{a}^{i}\right)- \dot{\nu}_{a}\left(t_{a}^{i-1}\right)\right\vert%
\right) \\
&&\leq c T_\Theta\#\Sigma_{t_1,t_2}(1+c\exp(c\tau_\Theta))\leq
\kappa\tau_\Theta(1+\kappa\exp(\kappa\tau_\Theta))(t_2-t_1+2T_\Theta).
\end{eqnarray*}
\end{proof}

\begin{lemma}
\label{discretizz} Let $(\Theta _{h})_{h\in {\mathbb{N}}}$ be a sequence of
time sets such that $T_{\Theta _{h}}\rightarrow 0$ as $h\rightarrow +\infty $%
. Let $((v_{h},z_{h}))_{h\in {\mathbb{N}}}\subset H^{1}(I,{{\mathbb{R}}%
^{D\times M}})$ converge strongly to some $(v,z)$, then 
\begin{equation}
\lim_{h}\mathcal{A}_{\mathcal{T},\Theta _{h}}(v_{h},z_{h})=\mathcal{A}_{%
\mathcal{T}}(v,z),  \label{actcont}
\end{equation}%
and for every $(\varphi _{h},\phi _{h})\in \mathcal{Y}_{\Theta _{h}}\cap
H_{0}^{1}(I,{{\mathbb{R}}^{D\times M}})$ such that $((\varphi _{h},\phi
_{h}))_{h\in {\mathbb{N}}}$ converges strongly to some $(\varphi ,\phi )\in
H_{0}^{1}(I,{{\mathbb{R}}^{D\times M}})$ 
\begin{equation}
\lim_{h}\delta \mathcal{A}_{\mathcal{T},\Theta _{h}}(v_{h},z_{h})[\varphi
_{h},\phi _{h}]=\delta \mathcal{A}_{\mathcal{T}}(v,z)[\varphi ,\phi ].
\label{firstvar}
\end{equation}%
Moreover, it holds 
\begin{equation}
\lim_{h}\mathcal{D}_{\mathcal{T},\Theta _{h}}(\dot{z}_{h})[\phi _{h}]=%
\mathcal{D}_{\mathcal{T}}(\dot{z})[\phi ].  \label{dissdiscrconv}
\end{equation}
\end{lemma}

\begin{proof}
The convergence property stated in \eqref{actcont} follows easily from the
strong convergence of $((v_h,z_h))_{h\in{\mathbb{N}}}$ in $H^{1}$ and the
regularity conditions of the integral densities of $V$ assumed in (A1)-(A3).

Define $\psi_h:=(v_h,z_h)$, $\psi:=(v,z)$, $\Phi_h=(\varphi_h,\phi_h)$, $%
\Phi:=(\varphi,\phi)$. Let $(\delta_j)$ be a positive vanishing sequence. By
taking into account \eqref{actcont}, for every $j\in{\mathbb{N}}$ we get 
\begin{equation*}
\lim_h\left(\frac{\mathcal{A}_{\mathcal{T},\Theta_h}(\psi_h+\delta_j\Phi_h)- 
\mathcal{A}_{\mathcal{T},\Theta_h}(\psi_h)}{\delta_j}- \frac{\mathcal{A}_{%
\mathcal{T}}(\psi+\delta_j\Phi)- \mathcal{A}_{\mathcal{T}}(\psi)}{\delta_j}%
\right)=0.
\end{equation*}
Hence, with fixed any subsequence $(h_i)$, by a diagonal argument we find a
further subsequence $(h_{i_j})$ such that 
\begin{equation}  \label{diagonal}
\lim_j\left(\frac{\mathcal{A}_{\mathcal{T},\Theta_{h_{i_j}}}
(\psi_{h_{i_j}}+\delta_j\Phi_{h_{i_j}})- \mathcal{A}_{\mathcal{T}%
,\Theta_{h_{i_j}}}(\psi_{h_{i_j}})}{\delta_j}- \frac{\mathcal{A}_{\mathcal{T}%
}(\psi+\delta_j\Phi)- \mathcal{A}_{\mathcal{T}}(\psi)}{\delta_j}\right)=0.
\end{equation}
Notice that by definition of first variation 
\begin{equation}  \label{diagonal1}
\delta\mathcal{A}_{\mathcal{T}}(\psi)[\Phi]= \lim_j\frac{\mathcal{A}_{%
\mathcal{T}}(\psi+\delta_j\Phi)- \mathcal{A}_{\mathcal{T}}(\psi)}{\delta_j}.
\end{equation}
Moreover, for every $h$ the action $\mathcal{A}_{\mathcal{T},\Theta_h}$ is a 
$C^{1}$ functional over $\mathcal{Y}_{\Theta_h}$ by the regularity
assumptions (A1)-(A3) on the densities of the potential $V$. Thus, for some $%
\varepsilon_j\in(0,\delta_j)$ we have 
\begin{equation}  \label{diagonal2}
\frac{\mathcal{A}_{\mathcal{T},\Theta_{h_{i_j}}}(\psi_{h_{i_j}}+
\delta_j\Phi_{h_{i_j}})- \mathcal{A}_{\mathcal{T},\Theta_{h_{i_j}}}(%
\psi_{h_{i_j}})}{\delta_j}= \delta\mathcal{A}_{\mathcal{T}%
,\Theta_{h_{i_j}}}(\psi_{h_{i_j}}
+\varepsilon_j\Phi_{h_{i_j}})[\Phi_{h_{i_j}}].
\end{equation}
Actually, the same regularity assumptions imply that 
\begin{equation}  \label{diagonal3}
\lim_j\left(\delta\mathcal{A}_{\mathcal{T},\Theta_{h_{i_j}}}(\psi_{h_{i_j}}
+\varepsilon_j\Phi_{h_{i_j}})[\Phi_{h_{i_j}}]- \delta\mathcal{A}_{\mathcal{T}%
,\Theta_{h_{i_j}}}(\psi_{h_{i_j}}) [\Phi_{h_{i_j}}]\right)=0.
\end{equation}
Thus, by \eqref{diagonal}-\eqref{diagonal3} we infer \eqref{firstvar} for
the subsequence $(h_{i_j})$. Urysohn property justifies \eqref{firstvar} for
the whole sequence.

To prove \eqref{dissdiscrconv}, note that the semi-discretized dissipation $%
\mathcal{D}_{\mathcal{T}}$ is continuous along sequences strongly converging
in $H^{1}$, so that 
\begin{equation*}
\lim_{h}\mathcal{D}_{\mathcal{T}}(\dot{z}_{h})[\phi _{h}]=\mathcal{D}_{%
\mathcal{T}}(z)[\phi ].
\end{equation*}%
To obtain \eqref{dissdiscrconv} we need only to prove 
\begin{equation*}
\lim_{h}\left( \mathcal{D}_{\mathcal{T},\Theta _{h}}(\dot{z}_{h})[\phi _{h}]-%
\mathcal{D}_{\mathcal{T}}(\dot{z}_{h})[\phi _{h}]\right) =0.
\end{equation*}%
The limit above 
is a consequence of \eqref{diffdiss} (since $T_{\Theta _{h}}\rightarrow 0$
as $h\rightarrow +\infty $) and the strong convergences in $H^{1}$ of $%
(z_{h})_{h\in \mathbb{N}}$ and $(\phi _{h})_{h\in \mathbb{N}}$ to $z$ and $%
\phi $, respectively.
\end{proof}

Proposition~\ref{stime} and Lemma~\ref{discretizz} are necessary tools for
the proof of Theorem~\ref{main}.

\begin{proof}[Proof of Theorem~\protect\ref{main}]
According to \eqref{incondbound} we may fix $\tau _{0}>0$ such that for
every $h$ 
\begin{equation*}
\tau _{\Theta _{h}}+\left\vert u_{h}(t_{0})\right\vert +\left\vert \nu
_{h}(t_{0})\right\vert +\left\vert \dot{u}_{h}(t_{0})\right\vert +\left\vert 
\dot{\nu}_{h}(t_{0})\right\vert \leq \tau _{0}.
\end{equation*}%
Denote by $(u_{h},\nu _{h})$ the solution to equations \eqref{Discr-dAL1}
and \eqref{Discr-dAL2}, associated with $\Theta _{h}$.

By Proposition~\ref{stime}, we infer that $((\dot{u}_h,\dot{\nu}_h))_{h\in{%
\mathbb{N}}}$ is bounded in $BV(I,{{\mathbb{R}}^{D\times M}})$ and by the $%
BV $ compactness and embedding theorems (see \cite[Theorem 3.23]{AFP}, \cite[%
Corollary 3.49]{AFP}), we deduce the existence of functions $(u,\nu)\in
W^{1,\infty}(\overline{I},{{\mathbb{R}}^{D\times M}})$ such that (up to a
sub-sequence) $((u_h,\nu_h))_{h\in{\mathbb{N}}}$ converges to $(u,\nu)$
uniformly on $\overline{I}$, and $((\dot{u}_h,\dot{\nu}_h))_{h\in{\mathbb{N}}%
}$ converges to $(\dot{u},\dot{\nu})$ in $L^p(I,{{\mathbb{R}}^{D\times M}})$
for every $p\in[1,+\infty)$.

Actually $(u,\nu )\in W^{2,\infty }(I,{{\mathbb{R}}^{D\times M}})$ as
observed for the one-dimensional case in \cite[Theorem 4.7]{MO}. We repeat
here the proof for the sake of completeness. According to \eqref{stimaBV3},
for every node $a\in {\mathcal{T}}$ and for every $t_{1},t_{2}\in I$ we have 
\begin{eqnarray}  \label{stimaBV5}
\lefteqn{|(\dot{u}_{h})_{a}(t_{2})-(\dot{u}_{h})_{a}(t_{1})|+|(\dot{\nu}%
_{h})_{a}(t_{2})-(\dot{\nu}_{h})_{a}(t_{1})|\notag} \\
&&\leq\kappa\tau _{0}\left(1+\exp\left(\kappa\tau _{0}\right) \right)
(t_{2}-t_{1}+2T_{\Theta _{h}}).
\end{eqnarray}%
Fix a standard mollifier $\psi \in C_{0}^{\infty }(-1,1)$, with $\psi \geq 0$
and $\int_{-1}^{1}\psi \,dx=1$ and set $\psi _{\delta }(x)=\delta ^{-1}\psi
(x/\delta )$. Let $(u_{h}^{\delta },\nu _{h}^{\delta })=(u_{h}\ast \psi
_{\delta },\nu _{h}\ast \psi _{\delta })$, it follows from \eqref{stimaBV5} 
\begin{eqnarray*}
\lefteqn{|(\dot{u}_{h})_{a}^{\delta }(t_{2})-(\dot{u}_{h})_{a}^{\delta
}(t_{1})|+|(\dot{\nu}_{h})_{a}^{\delta }(t_{2})-(\dot{\nu}_{h})_{a}^{\delta
}(t_{1})|} \\
&&\leq\kappa\tau_{0}\left( 1+\exp(\kappa\tau _{0})\right)
(t_{2}-t_{1}+2T_{\Theta_{h}})+c\delta .
\end{eqnarray*}%
By taking into account Lebesgue point theorem for $(\dot{u}_{a},{\dot{\nu}}%
_{a})$ and passing first to the limit as $h\rightarrow +\infty $ (note that
by assumption $T_{\Theta _{h}}\rightarrow 0$ as $h\rightarrow +\infty $) and
then as $\delta \rightarrow 0$ in the inequality above, we conclude that 
\begin{equation*}
|\dot{u}_{a}(t_{2})-\dot{u}_{a}(t_{1})|+|{\dot{\nu}}_{a}(t_{2})-{\dot{\nu}}%
_{a}(t_{1})|\leq\kappa\tau _{0}(1+\exp(\kappa\tau _{0}))(t_{2}-t_{1}).
\end{equation*}%
Eventually, we show that $(u,\nu )$ solves \eqref{ddAL} 
by passing to the limit in \eqref{dALdiscr}. Take test functions $(\varphi
,\phi )\in H_{0}^{1}\left( I,{{\mathbb{R}}^{D\times M}}\right) $, and let $%
(\varphi _{h},\phi _{h})\in \mathcal{Y}_{\Theta _{h}}\cap H_{0}^{1}(I,{{%
\mathbb{R}}^{D\times M}})$ be a sequence of piecewise affine test functions
converging to $(\varphi ,\phi )$ strongly in $H^{1}\left( I,{{\mathbb{R}}%
^{D\times M}}\right) $ (see for instance \cite[Lemma 4.3]{MO}). Then the
discretized d'Alembert-Lagrange-type principle~\eqref{dALdiscr} is
satisfied, namely 
\begin{equation*}
\delta \mathcal{A}_{\mathcal{T},\Theta _{h}}(u_{h},\nu _{h})[\varphi
_{h},\phi _{h}]+\mathcal{D}_{\mathcal{T},\Theta _{h}}({\dot{\nu}}_{h})[\phi
_{h}]=0.
\end{equation*}%
The conclusion then follows by Lemma~\ref{discretizz} and the strong
convergence in $H^{1}$ of all the relevant sequences.
\end{proof}

\begin{remark}
\label{local} A standard diagonalization argument implies that we can allow
for intervals $I=(t_0,+\infty)$ provided that all the statements in Theorem~%
\ref{main} are intended in a local sense.
\end{remark}

\begin{remark}
\label{general} The methods developed to prove Theorem~\ref{main} can be
refined to deal with kinetic co-energies $\chi$ of the form 
\begin{equation*}
\chi(\nu,{\dot{\nu}})=\frac 12 \Omega{\dot{\nu}}\cdot{\dot{\nu}}
\end{equation*}
where $\Omega$ is a positive definite symmetric element of $Hom(T_\nu%
\mathcal{M},T_\nu^*\mathcal{M})$. The idea is to choose for the fields $u$, $%
\nu$ different shape functions: for $u$ the family $\{\mathcal{N}_a\}_{a\in%
\mathcal{T}}$ as above, and for $\nu$ a family $\{\mathcal{N}%
_a^\prime\}_{a\in\mathcal{T}}$ in such a way that the mass matrix associated
with the spatial discretization of $\chi$ is in diagonal form.
\end{remark}

\section{Synchronous Variational Integrators}

\label{VIs}

Although the choices $\frac{1}{2}\overline{\rho }\dot{\nu}^{\flat }\cdot 
\dot{\nu}$ and $\frac{1}{2}\Omega {\dot{\nu}}\cdot {\dot{\nu}}$ as
expressions of the substructural kinetic energy cover a wide set of special
cases, there are circumstances in which more complicated expressions arise
(see for example the dynamics of liquids with a dense distribution of
bubbles \cite{C89}). For this reason the analysis of the dicrete schemes in
which the substructural kinetic co-energy is left unspecified has its own
importance. In this case we prove the convergence of the discrete dynamics
only for synchronous variational integrators, i.e. when $\Theta _{K}$
coincide with $\Theta $ for every $K\in \mathcal{T}$.

More precisely, we assume that $\chi$ is any function with the following
properties 

\begin{itemize}
\item[(A4)] $\chi\in C^{2}(\mathbb{R}^{k\times k},[0,+\infty))$ has \emph{%
quadratic growth}, i.e. there exists a constant $\Xi>0$ such that for all $%
\nu,\zeta\in{{\mathbb{R}}^k}$ 
\begin{equation*}
|\nabla^{2}\chi(\nu,\zeta)|\leq \Xi.
\end{equation*}

\item[(A5)] $\chi (\nu ,\cdot )$ is \emph{uniformly convex} with respect to $%
\nu $, i.e. there exists a constant $\gamma >0$ such that for all $\nu
,\zeta ,\xi \in {{\mathbb{R}}^{k}}$ 
\begin{equation*}
\partial _{{\dot{\nu}}{\dot{\nu}}}^{2}\chi (\nu ,\zeta )\cdot \left( \xi
\otimes \xi \right) \geq \gamma |\xi |^{2}.
\end{equation*}
\end{itemize}

The spatial discretization follows the same lines indicated in Subsection~%
\ref{spacetime}. In the sequel we use the same notations taken there, the
only difference being that the family of shape functions $\{\mathcal{N}%
_{a}\}_{a\in \mathcal{T}}$ is chosen to form an orthonormal system in $L^{2}(%
\mathcal{B})$ (see Remark~\ref{notorthon}). A discrete version in space of
the d'Alembert-Lagrange-type principle~\eqref{ddALcont} arises. We then get 
\begin{eqnarray}
\lefteqn{\int_{I}m_{a}\dot{u}_{a}\left( t\right) \cdot \dot{\varphi}%
_{a}(t)\,dt=\int_{I}\left( \int_{\mathcal{B}}\partial _{F}e\left( (\nabla
u,\nu ,\nabla \nu )(x,t)\right) \nabla \mathcal{N}_{a}(x)\,dx\right) \cdot
\varphi _{a}(t)\,dt}  \notag  \label{dALL1s} \\
&&+\int_{I}\left( \int_{\mathcal{B}}\partial _{u}w\left( (u,\nu
)(x,t)\right) \mathcal{N}_{a}(x)dx-\int_{\partial \mathcal{B}_{t}}\mathsf{t}%
(x)\mathcal{N}_{a}(x)\,d\mathcal{H}^{d-1}(x)\right) \cdot \varphi
_{a}(t)\,dt,
\end{eqnarray}%
and 
\begin{eqnarray}
\lefteqn{\int_{I}\left( \int_{\mathcal{B}}\partial _{\dot{\nu}}\chi \left(
(\nu ,\dot{\nu})(x,t)\right) \mathcal{N}_{a}(x)\,dx\right) \cdot \dot{\phi}%
_{a}(t)\,dt}  \notag  \label{dALL2s} \\
&&+\int_{I}\left( \int_{\mathcal{B}}\partial _{\nu }\chi \left( (\nu ,{\dot{%
\nu}})(x,t)\right) \mathcal{N}_{a}(x)\,dx\right) \cdot \phi _{a}(t)\,dt+\eta
\int_{I}{\dot{\nu}}_{a}(t)\cdot \phi _{a}(t)\,dt  \notag \\
&=&\int_{I}\left( \int_{\mathcal{B}}\partial _{N}e\left( (\nabla u,\nu
,\nabla \nu )(x,t)\right) \nabla \mathcal{N}_{a}(x)\,dx\right) \cdot \phi
_{a}(t)\,dt  \notag \\
&&+\int_{I}\left( \int_{\mathcal{B}}\left( \partial _{\nu }e\left( (\nabla
u,\nu ,\nabla \nu )(x,t)\right) +\partial _{\nu }w\left( (u,\nu
)(x,t)\right) \right) \mathcal{N}_{a}(x)\,dx\right) \cdot \phi _{a}(t)\,dt,
\end{eqnarray}%
where the test functions satisfy $\varphi _{a}\in H_{0}^{1}(I,{{\mathbb{R}}%
^{d}})$ and $\varphi _{a}\equiv 0$ if $a\in \partial \mathcal{B}_{u}$, and $%
\phi _{a}\in H_{0}^{1}(I,{{\mathbb{R}}^{k}})$ and $\phi _{a}\equiv 0$ if $%
a\in \partial \mathcal{B}_{\nu }$.

As regards the time discretization, we fix a discrete time set $\Theta
=\{t_{i}\}_{i=1,\ldots ,N_{\Theta }}$ for $I$, with time size $h_{\Theta
}=\max_{\Theta }(t_{i+1}-t_{i})$, and consider the subspace $\mathcal{Y}%
_{\Theta }$ of $H^{1}(I,{{\mathbb{R}}^{D\times M}})$ of continuous functions
which are affine on each interval $(t_{i},t_{i+1})$ and such that $%
u_{a}\equiv 0$ if $a\in \partial \mathcal{B}_{u}$ and $\nu _{a}\equiv 0$ if $%
a\in \partial \mathcal{B}_{\nu }$. We then define the discrete action as 
\begin{equation*}
\mathcal{A}_{\mathcal{T},\Theta }\left( u,\nu \right) :=%
\begin{cases}
\displaystyle{\sum_{i=0}^{N_{\Theta }-1}(t_{i+1}-t_{i})\mathcal{A}_{\mathcal{%
T},\Theta }^{i}\left( u,\nu \right) } & \text{ if }(u,\nu )\in \mathcal{Y}%
_{\Theta }\cr+\infty & \text{ if }(u,\nu )\in \mathcal{Y}\backslash \mathcal{%
Y}_{\Theta },%
\end{cases}%
\end{equation*}%
where 
\begin{equation*}
\mathcal{A}_{\mathcal{T},\Theta }^{i}\left( u,\nu \right) :=\sum_{a\in 
\mathcal{T}}\frac{1}{2}|\dot{u}_{a}(t_{i})|^{2}+\int_{\mathcal{B}}\chi ((\nu
,{\dot{\nu}})(x,t_{i}))\,dx-V((u,\nu )(t_{i})),
\end{equation*}%
and the potential $V$ is defined in \eqref{potential}. Notice that in this
case $\Theta _{K}\equiv \Theta $ for every $K\in \mathcal{T}$, and thus $%
T_{\Theta }=h_{\Theta }$, $\tau _{\Theta }=1$. We denote by $\tau _{\Theta
}^{\prime }$ the ratio 
\begin{equation}
\tau _{\Theta }^{\prime }:=\frac{\max_{\Theta }\left( t_{i+1}-t_{i}\right) }{%
\min_{\Theta }\left( t_{i+1}-t_{i}\right) }.  \label{MTetas}
\end{equation}%
Note that we have chosen the approximation 
\begin{eqnarray*}
\lefteqn{\int_{t_{i}}^{t_{i+1}}\left( \int_{\mathcal{B}}\chi ((\nu ,\dot{\nu}%
)(x,t))\,dx-V\left( (u,\nu )(t)\right) \right) dt} \\
&\approx &(t_{i+1}-t_{i})\left( \int_{\mathcal{B}}\chi ((\nu ,{\dot{\nu}}%
)(x,t_{i}))\,dx-V((u,\nu )(t_{i}))\right) .
\end{eqnarray*}%
In this setting, the d'Alembert-Lagrange-type discrete principle is then
given by 
\begin{equation}
\delta \mathcal{A}_{\mathcal{T},\Theta }(u,\nu )[\varphi ,\phi ]+\mathcal{D}%
_{\mathcal{T},\Theta }({\dot{\nu}})[\phi ]=0,  \label{dALLsynch}
\end{equation}%
where $(\varphi ,\phi )\in \mathcal{Y}_{\Theta }\cap H_{0}^{1}(I,{{\mathbb{R}%
}^{D\times M}})$ and $\mathcal{D}_{\mathcal{T},\Theta }$ is defined as in %
\eqref{dissdiscr}, namely 
\begin{equation*}
\mathcal{D}_{\mathcal{T},\Theta }({\dot{\nu}})[\phi ]:=\eta \sum_{a\in 
\mathcal{T}}\sum_{i=1}^{N_{\Theta }}(\nu _{a}(t_{i})-\nu _{a}(t_{i-1}))\cdot
\phi _{a}(t_{i-1}).
\end{equation*}%
The main result of the section reads as follows.

\begin{theorem}
\label{main2} Assume (A1)-(A5). Let $(\Theta _{h})_{h\in {\mathbb{N}}}$ be a
sequence of time sets for a bounded time interval $I=(t_{0},t_{f})$ such
that $T_{\Theta _{h}}\rightarrow 0$ as $h\rightarrow +\infty $. Let also $%
(u_{h}(t_{0}),\nu _{h}(t_{0}))$ and $(\dot{u}_{h}(t_{0}),\dot{\nu}%
_{h}(t_{0}))$ be initial conditions satisfying 
\begin{equation}
\sup_{h}\left( \tau _{\Theta _{h}}^{\prime }+\left\vert
u_{h}(t_{0})\right\vert +\left\vert \nu _{h}(t_{0})\right\vert +\left\vert 
\dot{u}_{h}(t_{0})\right\vert +\left\vert \dot{\nu}_{h}(t_{0})\right\vert
\right) <+\infty .  \label{incondbound2}
\end{equation}%
The discrete d'Alembert-Lagrange-type principle \eqref{dALLsynch} relative
to $\Theta _{h}$ has a unique solution $(u_{h},\nu _{h})$ for $h$
sufficiently large.

Moreover, there exists $(u,\nu)\in W^{2,\infty}(I,{{\mathbb{R}}^{D\times M}}%
) $ satisfying \eqref{dALL1s} and \eqref{dALL2s} such that, up to a
sub-sequence, $((u_h,\nu_h))_{h\in\mathbb{N}}$ converges to $(u,\nu)$
uniformly on $\overline{I}$, and $((\dot{u}_h,\dot{\nu}_h))_{h\in\mathbb{N}}$
converges to $(\dot{u},\dot{\nu})$ in $L^{p}(I,{{\mathbb{R}}^{D\times M}})$
for every $p\in[1,+\infty)$.
\end{theorem}

The proof follows the same strategy developed for the proof of Theorem~\ref%
{main}, once all the preparatory steps have been accomplished. Thus we will
not give the details of the proof of Theorem~\ref{main2}. Instead, we will
focus our attention to show existence and uniqueness of discrete solutions
to \eqref{dALLsynch} and the $BV$ estimates on their velocities. In doing
that the main ingredient that we will exploit is the uniform convexity of $%
\chi(\nu,\cdot)$ assumed in (A5).

In the following lemma we collect all the properties of $\chi$ which will be
useful in the sequel. The proof of the lemma is standard and then omitted.

\begin{lemma}
\label{tools} Assume (A4) and (A5), then the following properties hold true

\begin{itemize}
\item[(1)] \emph{Uniform Strict Monotonicity of $\partial_{{\dot{\nu}}%
}\chi(\nu,\cdot)$}: For all $\nu,\zeta_1,\zeta_2\in{{\mathbb{R}}^k}$ 
\begin{equation*}
\gamma|\zeta_1-\zeta_2|^2\le \langle\partial_{{\dot{\nu}}}\chi(\nu,\zeta_1)-
\partial_{{\dot{\nu}}}\chi(\nu,\zeta_2),\zeta_1-\zeta_2\rangle.
\end{equation*}

\item[(2)] \emph{Uniform Lipschitz Continuity of $\partial_{{\dot{\nu}}%
}\chi(\cdot,{\dot{\nu}})$, $\partial_{\nu}\chi(\nu,\cdot)$}: For all $%
\nu_1,\nu_2,\zeta\in{{\mathbb{R}}^k}$ 
\begin{equation*}
|\partial_{{\dot{\nu}}}\chi(\nu_1,\zeta)-\partial_{{\dot{\nu}}}
\chi(\nu_2,\zeta)|\leq\Xi|\nu_1-\nu_2|,
\end{equation*}
and for all $\nu,\zeta_1,\zeta_2\in{{\mathbb{R}}^k}$ 
\begin{equation*}
|\partial_{\nu}\chi(\nu,\zeta_1)-\partial_{\nu}\chi(\nu,\zeta_2)|
\leq\Xi|\zeta_1-\zeta_2|.
\end{equation*}

\item[(3)] \emph{Growth conditions of $\chi$, $\partial_{{\dot{\nu}}}\chi$, $%
\partial_{\nu}\chi$}: There exists a constant $\Xi_2>0$ such that for all $%
\nu,\zeta\in{{\mathbb{R}}^k}$ 
\begin{eqnarray*}
&&|\partial_\nu\chi(\nu,\zeta)|+|\partial_{{\dot{\nu}}}\chi(\nu,\zeta)|
\leq\Xi_2(1+|\nu|+|\zeta|), \\
&&\gamma|\zeta|^2-\Xi_2(1+|\nu|^2)\leq
\chi(\nu,\zeta)\leq\Xi_2(1+|\nu|^2+|\zeta|^2).
\end{eqnarray*}
\end{itemize}
\end{lemma}

\subsection{Convergence in time}

As a first step we write an equivalent form of \eqref{dALLsynch} which can
be obtained by varying only one node at each time $t_i\in\Theta$.

\begin{lemma}
\label{ddALexplsynch} Let $(u,\nu )\in \mathcal{Y}_{\Theta }$. $(u,\nu )$
solves \eqref{dALLsynch} if and only if given any node $a\in \mathcal{T}%
\backslash (\partial \mathcal{B}_{u}\cup \partial \mathcal{B}_{\nu })$ and
any time $t_{i}\in \Theta \cap (t_{0},t_{f}]$, the following balance
equations hold: 
\begin{eqnarray}
\lefteqn{m_{a}(\dot{u}_{a}\left( t_{i-1}\right) -\dot{u}_{a}\left(
t_{i}\right) )=(t_{i+1}-t_{i})\int_{\mathcal{B}}\partial _{F}e\left( (\nabla
u,\nu ,\nabla \nu )(x,t_{i})\right) \nabla \mathcal{N}_{a}(x)\,dx}  \notag
\label{Discr-dAL1s} \\
&&+(t_{i+1}-t_{i})\left( \int_{\mathcal{B}}\partial _{u}w\left( (u,\nu
)(x,t_{i})\right) \mathcal{N}_{a}(x)\,dx-\int_{\partial \mathcal{B}_{t}}%
\mathsf{t}(x)\mathcal{N}_{a}(x)\,d\mathcal{H}^{d-1}\right) ,
\end{eqnarray}%
and 
\begin{eqnarray}
\lefteqn{\int_{\mathcal{B}}\left( \partial _{\dot{\nu}}\chi \left( (\nu ,%
\dot{\nu})(x,t_{i-1})\right) -\partial _{\dot{\nu}}\chi \left( (\nu ,\dot{\nu%
})(x,t_{i})\right) \right) \mathcal{N}_{a}(x)\,dx}  \notag
\label{Discr-dAL2s} \\
&&+(t_{i+1}-t_{i})\int_{\mathcal{B}}\partial _{\nu }\chi \left( (\nu ,{\dot{%
\nu}})(x,t_{i})\right) \mathcal{N}_{a}(x)\,dx+\eta (\nu _{a}(t_{i+1})-\nu
_{a}(t_{i}))  \notag \\
&=&(t_{i+1}-t_{i})\int_{\mathcal{B}}\partial _{N}e\left( (\nabla u,\nu
,\nabla \nu )(x,t_{i})\right) \nabla \mathcal{N}_{a}(x)\,dx  \notag \\
&&+(t_{i+1}-t_{i})\int_{\mathcal{B}}\left( \partial _{\nu }e\left( (\nabla
u,\nu ,\nabla \nu )(x,t_{i})\right) +\partial _{\nu }w\left( (u,\nu
)(x,t_{i})\right) \right) \mathcal{N}_{a}(x)\,dx.
\end{eqnarray}%
In case $a\in \partial \mathcal{B}_{\nu }\backslash \partial \mathcal{B}_{u}$
only equation \eqref{Discr-dAL1s} has to be satisfied, while if $a\in
\partial \mathcal{B}_{u}\backslash \partial \mathcal{B}_{\nu }$ the sole %
\eqref{Discr-dAL2s} has to be satisfied.
\end{lemma}

If for a fixed time $t_{i}\in \Theta \cap \lbrack t_{0},t_{f})$ a solution
to \eqref{Discr-dAL1s} and \eqref{Discr-dAL2s} exists for every node in $%
\mathcal{T}$, a unique solution at time $t_{i+1}$ is determined. An
induction argument provides the conclusion once the initial conditions are
specified.

Consider the function $\Psi :\mathbb{R}^{k\times \#(\mathcal{T}\setminus
\partial \mathcal{B}_{\nu })}\rightarrow \mathbb{R}^{k\times \#(\mathcal{T}%
\setminus \partial \mathcal{B}_{\nu })}$, $\Psi =(\Psi _{a})_{a\in \mathcal{T%
}\setminus \partial \mathcal{B}_{\nu }}$, where for every node $a\in 
\mathcal{T}\setminus \partial \mathcal{B}_{\nu }$ 
\begin{eqnarray}
\lefteqn{\Psi _{a}(\nu ):=\int_{\mathcal{B}}\partial _{\dot{\nu}}\chi \left(
\nu (x,t_{i}),\sum_{\alpha \in \mathcal{T}}\mathcal{N}_{\alpha }(x)\frac{\nu
_{\alpha }-\nu _{\alpha }(t_{i})}{t_{i+1}-t_{i}}\right) \mathcal{N}%
_{a}(x)\,dx}  \notag  \label{psia} \\
&&-\int_{\mathcal{B}}\partial _{{\dot{\nu}}}\chi \left( (\nu ,{\dot{\nu}}%
)(x,t_{i-1})\right) \mathcal{N}_{a}(x)\,dx+\eta (\nu _{a}-\nu _{a}(t_{i})) 
\notag \\
&&+(t_{i+1}-t_{i})\int_{\mathcal{B}}\partial _{\nu }\chi \left( \nu
(x,t_{i}),\sum_{\alpha \in \mathcal{T}}\mathcal{N}_{\alpha }(x)\frac{\nu
_{\alpha }-\nu _{\alpha }(t_{i})}{t_{i+1}-t_{i}}\right) \mathcal{N}%
_{a}(x)\,dx.
\end{eqnarray}%
Notice that for every node $a\in \mathcal{T}\setminus \partial \mathcal{B}%
_{\nu }$ 
equation~\eqref{Discr-dAL2s} becomes%
\begin{eqnarray}
\lefteqn{\Psi _{a}\left( (\nu _{\alpha }(t_{i+1}))_{\alpha \in \mathcal{T}%
\setminus \partial \mathcal{B}_{\nu }}\right) =(t_{i+1}-t_{i})\int_{\mathcal{%
B}}\partial _{N}e\left( (\nabla u,\nu ,\nabla \nu )(x,t_{i})\right) \nabla 
\mathcal{N}_{a}(x)\,dx}  \notag  \label{uguagl} \\
&&+(t_{i+1}-t_{i})\int_{\mathcal{B}}\left( \partial _{\nu }e\left( (\nabla
u,\nu ,\nabla \nu )(x,t_{i})\right) +\partial _{\nu }w\left( (u,\nu
)(x,t_{i})\right) \right) \mathcal{N}_{a}(x)\,dx.
\end{eqnarray}%
Recall that $\nu _{a}(t)\equiv 0$ for every time $t\in I$ if $a\in \partial 
\mathcal{B}_{\nu }$.

\begin{lemma}
\label{invertib} There exists $T_0=T_0(\gamma,\Xi)>0$ such that, for any
time set $\Theta$ with $T_\Theta\in(0,T_0)$, the map $\Psi$ is a bijection.
Moreover, for every $\nu,\mu \in{\mathbb{R}}^{k\times\#(\mathcal{T}%
\setminus\partial\mathcal{B}_\nu)}$ 
\begin{equation}  \label{invert}
\frac\gamma 2 \sum_{a\in\mathcal{T}\setminus\partial\mathcal{B}_\nu}
\left|\nu_a-\mu_a\right|^2\leq\langle\Psi(\nu)-\Psi(\mu),\nu-\mu\rangle.
\end{equation}
\end{lemma}

\begin{proof}
Let us show the injectivity of $\Psi $. Given $\nu $ and $\mu $ in ${\mathbb{%
R}}^{k\times \#(\mathcal{T}\setminus \partial \mathcal{B}_{\nu })}$, the
uniform strict monotonicity of $\partial _{{\dot{\nu}}}\chi (\nu ,\cdot )$
(see (1) in Lemma~\ref{tools}), the uniform Lipschitz continuity of $%
\partial _{\nu }\chi (\cdot ,{\dot{\nu}})$ (see (2) in Lemma~\ref{tools}),
and the orthonormality of $\{\mathcal{N}_{a}\}_{a\in {\mathcal{T}}}$ in $%
L^{2}({\mathcal{B}})$ imply 
\begin{eqnarray*}
\lefteqn{\langle \Psi (\nu )-\Psi (\mu ),\nu -\mu \rangle =\sum_{a\in 
\mathcal{T}\setminus \partial \mathcal{B}_{\nu }}\langle \Psi _{a}(\nu
)-\Psi _{a}(\mu ),\nu _{a}-\mu _{a}\rangle } \\
&\geq &\eta \sum_{a\in \mathcal{T}\setminus \partial \mathcal{B}_{\nu }}|\nu
_{a}-\mu _{a}|^{2}+\left( \frac{\gamma }{t_{i+1}-t_{i}}-\Xi \right)
\sum_{a\in \mathcal{T}\setminus \partial \mathcal{B}_{\nu }}\left\vert \nu
_{a}-\mu _{a}\right\vert ^{2} \\
&\geq &\left( \frac{\gamma }{t_{i+1}-t_{i}}-\Xi \right) \sum_{a\in \mathcal{T%
}\setminus \partial \mathcal{B}_{\nu }}\left\vert \nu _{a}-\mu
_{a}\right\vert ^{2}.
\end{eqnarray*}%
If $T_{0}\leq 2\gamma /(2\Xi +\gamma )$, the latter equation implies
inequality \eqref{invert} for $T_{\Theta }\in (0,T_{0})$. The injectivity of 
$\Psi $ is then established.

Eventually, the surjectivity of $\Psi$ follows at once from \eqref{invert}
by arguing componentwise.
\end{proof}

Existence and uniqueness of solutions to \eqref{Discr-dAL1s} and %
\eqref{Discr-dAL2s} can be shown.

\begin{proposition}
\label{existence} If $T_{0}=T_{0}(\gamma ,\Xi )>0$ is the constant in Lemma~%
\ref{invertib}, for every time set $\Theta $ with size $T_{\Theta }\in
(0,T_{0})$, a unique solution to equations \eqref{Discr-dAL1s} and %
\eqref{Discr-dAL2s} exists in $\mathcal{Y}_{\Theta }$ once initial
conditions $(u(t_{0}),\nu (t_{0}))$, $(\dot{u}(t_{0}),{\dot{\nu}}(t_{0}))$
are given.
\end{proposition}

\begin{proof}
By taking into account \eqref{uguagl} and the invertibility of $\Psi$ proved
in Lemma~\ref{invertib} (for $T_\Theta\in(0,T_0)$), \eqref{Discr-dAL1s} and %
\eqref{Discr-dAL2s} define inductively a unique piecewise affine trajectory $%
(u,\nu)\in \mathcal{Y}_\Theta$ once the initial conditions are specified.
\end{proof}

It is clear that in this setting the variational approach based on $\Gamma $%
-convergence proposed in \cite{MO} cannot be pursued. In the conservative
case a more conventional approach based on the repeated use of Gronwall
inequality has been developed in \cite{LMOW2}.

The following proposition is the analogue of Proposition~\ref{stime}. Let us
point out that 
the $BV$ estimates proven in the sequel are instrumental. Indeed, they
enable us to pass to the limit directly into the discrete equations %
\eqref{Discr-dAL1s} and \eqref{Discr-dAL2s} and to prove that their
solutions have cluster points solving \eqref{dALL1s} and \eqref{dALL2s}.

\begin{proposition}
\label{stimes} There exists a constant $T_{1}=T_{1}(\gamma ,\Xi ,\eta
,\lambda ,\Lambda ,\mathcal{T})>0$ such that given initial conditions $%
\left( u(t_{0}),\nu (t_{0})\right) $ and $\left( \dot{u}(t_{0}),\dot{\nu}%
(t_{0})\right) $ for every entire time set $\Theta $ with $T_{\Theta }\in
(0,T_{1})$, the solution $(u,\nu )\in \mathcal{Y}_{\Theta }$ to equations %
\eqref{Discr-dAL1s} and \eqref{Discr-dAL2s} satisfies 
\begin{equation}
\Vert \dot{u}\Vert _{L^{\infty }\left( I,{{\mathbb{R}}^{D}}\right) }+\Vert 
\dot{\nu}\Vert _{L^{\infty }\left( I,{{\mathbb{R}}^{M}}\right) }\leq \kappa
\exp \left( \kappa \tau _{\Theta }^{\prime }\right) ,  \label{Prop1s}
\end{equation}%
for some constant $\kappa >0$ depending on the data of the problem and on
the initial conditions themselves.

Moreover, the functions $\dot{u}$, ${\dot{\nu}}$ have (pointwise) bounded
variation on $\overline{I}$ with 
\begin{equation}  \label{stimaBVs}
pV(\dot{u},[t_1,t_2])+pV(\dot{\nu},[t_1,t_2]) \leq\kappa
\tau_\Theta^\prime\left(1+\exp\left(\kappa\tau_\Theta^\prime\right)\right)
(t_2-t_1+2T_\Theta),
\end{equation}
for every interval $[t_1,t_2]\subseteq \overline{I}$.
\end{proposition}

\begin{proof}
Fix a time $t_{i}\in \Theta $. To estimate the velocity of $u$ at time $%
t_{i} $ in terms of the velocities of $u$ itself and $\nu $ at previous
times we follows the path indicated in Proposition~\ref{stime}. We use
equation \eqref{Discr-dAL1s} and, for every node $a\in \mathcal{T}$, we get 
\begin{equation}
\left\vert \dot{u}_{a}\left( t_{i}\right) -\dot{u}_{a}\left( t_{i-1}\right)
\right\vert \leq cT_{\Theta }\left( 1+\Vert \dot{u}\Vert _{L^{\infty }\left(
(t_{0},t_{i-1}),{{\mathbb{R}}^{D}}\right) }+\Vert \dot{\nu}\Vert _{L^{\infty
}\left( (t_{0},t_{i-1}),{{\mathbb{R}}^{M}}\right) }\right) .
\label{stimaLinf1s}
\end{equation}%
The latter inequality yields 
\begin{equation}
\left\Vert \dot{u}\right\Vert _{L^{\infty }\left( (t_{0},t_{i}),{{\mathbb{R}}%
^{D}}\right) }\leq cT_{\Theta }\left( 1+\left\Vert \dot{\nu}\right\Vert
_{L^{\infty }\left( (t_{0},t_{i-1}),{{\mathbb{R}}^{M}}\right) }\right)
+(1+cT_{\Theta })\left\Vert \dot{u}\right\Vert _{L^{\infty }\left(
(t_{0},t_{i-1}),{{\mathbb{R}}^{D}}\right) }.  \label{stimaLinf2s}
\end{equation}%
To prove the analogous estimate for ${\dot{\nu}}$ we use the equivalent form
of equation \eqref{Discr-dAL2s} written in \eqref{uguagl}. We rewrite the
left hand side in \eqref{uguagl} as 
\begin{equation*}
\Psi _{a}((\nu _{\alpha }(t_{i+1}))_{\alpha \in \mathcal{T}\setminus
\partial \mathcal{B}_{\nu }})=J_{a}^{1}+J_{a}^{2}+J_{a}^{3},
\end{equation*}%
where 
\begin{eqnarray*}
&&J_{a}^{1}:=\int_{\mathcal{B}}\left( \partial _{{\dot{\nu}}}\chi \left(
(\nu ,{\dot{\nu}})(x,t_{i})\right) -\partial _{{\dot{\nu}}}\chi \left( \nu
(x,t_{i}),{\dot{\nu}}(x,t_{i-1})\right) \right) \mathcal{N}_{a}(x)\,dx, \\
&&J_{a}^{2}:=\int_{\mathcal{B}}\left( \partial _{{\dot{\nu}}}\chi \left( \nu
(x,t_{i}),{\dot{\nu}}(x,t_{i-1})\right) -\partial _{{\dot{\nu}}}\chi \left(
(\nu ,{\dot{\nu}})(x,t_{i-1})\right) \right) \mathcal{N}_{a}(x)\,dx, \\
&&J_{a}^{3}:=-(t_{i+1}-t_{i})\int_{\mathcal{B}}\partial _{\nu }\chi \left(
(\nu ,{\dot{\nu}})(x,t_{i})\right) \mathcal{N}_{a}(x)\,dx+\eta (\nu
_{a}(t_{i+1})-\nu _{a}(t_{i})).
\end{eqnarray*}%
To evaluate the variation of ${\dot{\nu}}$ we consider the scalar product of 
$\Psi $ calculated in $(\nu _{\alpha }(t_{i+1}))_{\alpha \in \mathcal{T}%
\setminus \partial \mathcal{B}_{\nu }}$ with the vector $\overline{\nu }:=({%
\dot{\nu}}_{a}(t_{i})-{\dot{\nu}}_{a}(t_{i-1}))_{a\in \mathcal{T}\setminus
\partial \mathcal{B}_{\nu }}$, so that we get 
\begin{equation*}
\langle \Psi ((\nu _{\alpha }(t_{i+1}))_{\alpha \in \mathcal{T}\setminus
\partial \mathcal{B}_{\nu }}),\overline{\nu }\rangle
=\sum_{r=1,2,3}\sum_{a\in \mathcal{T}}\langle J_{a}^{r},{\dot{\nu}}%
_{a}(t_{i})-{\dot{\nu}}_{a}(t_{i-1})\rangle .
\end{equation*}%
Indeed, recall that ${\dot{\nu}}_{a}(t_{i})=0$ for every $i$ if $a\in
\partial \mathcal{B}_{\nu }$. We estimate separately each term above. First
by (1) in Lemma~\ref{tools} we have 
\begin{eqnarray}
&&\sum_{a\in \mathcal{T}}\langle J_{a}^{1},{\dot{\nu}}_{a}(t_{i})-{\dot{\nu}}%
_{a}(t_{i-1})\rangle =  \notag  \label{e1} \\
&&\int_{\mathcal{B}}\langle \partial _{{\dot{\nu}}}\chi \left( (\nu ,{\dot{%
\nu}})(x,t_{i})\right) -\partial _{{\dot{\nu}}}\chi \left( \nu (x,t_{i}),{%
\dot{\nu}}(x,t_{i-1})\right) ,{\dot{\nu}}(x,t_{i})-{\dot{\nu}}%
(x,t_{i-1})\rangle \,dx  \notag \\
&\geq &\gamma \int_{\mathcal{B}}|{\dot{\nu}}(x,t_{i})-{\dot{\nu}}%
(x,t_{i-1})|^{2}\,dx=\gamma \sum_{a\in \mathcal{T}}|{\dot{\nu}}_{a}(t_{i})-{%
\dot{\nu}}_{a}(t_{i-1})|^{2}=\gamma |\overline{\nu }|^{2}.
\end{eqnarray}%
Then, by (2) in Lemma~\ref{tools} and since $\nu (x,t_{i})-\nu (x,t_{i-1})={%
\dot{\nu}}(t_{i-1})(t_{i}-t_{i-1})$, by H\"{o}lder inequality we obtain 
\begin{eqnarray}
&&\sum_{a\in \mathcal{T}}\langle J_{a}^{2},{\dot{\nu}}_{a}(t_{i})-{\dot{\nu}}%
_{a}(t_{i-1})\rangle  \notag  \label{e2} \\
&=&\int_{\mathcal{B}}\langle \partial _{{\dot{\nu}}}\chi \left( \nu
(x,t_{i}),{\dot{\nu}}(x,t_{i-1})\right) -\partial _{{\dot{\nu}}}\chi \left(
(\nu ,{\dot{\nu}})(x,t_{i-1})\right) ,{\dot{\nu}}(x,t_{i})-{\dot{\nu}}%
(x,t_{i-1})\rangle \,dx  \notag \\
&\geq &-\Xi T_{\Theta }\Vert {\dot{\nu}}\Vert _{L^{\infty }((t_{0},t_{i-1}),{%
{\mathbb{R}}^{M}})}|\overline{\nu }|.
\end{eqnarray}%
Moreover, by (3) in Lemma~\ref{tools} we infer 
\begin{eqnarray}
&&\sum_{a\in \mathcal{T}}\langle J_{a}^{3},{\dot{\nu}}_{a}(t_{i})-{\dot{\nu}}%
_{a}(t_{i-1})\rangle =\eta (t_{i+1}-t_{i})\sum_{a\in \mathcal{T}}\langle {%
\dot{\nu}}_{a}(t_{i}),{\dot{\nu}}_{a}(t_{i})-{\dot{\nu}}_{a}(t_{i-1})\rangle
\notag  \label{e3} \\
&&-(t_{i+1}-t_{i})\int_{\mathcal{B}}\langle \partial _{\nu }\chi \left( (\nu
,{\dot{\nu}})(x,t_{i})\right) ,{\dot{\nu}}(x,t_{i})-{\dot{\nu}}%
(x,t_{i-1})\rangle \,dx  \notag \\
&\geq &-cT_{\Theta }(\eta +\Xi _{1})\left( 1+\Vert \nu \Vert _{L^{\infty
}((t_{0},t_{i}),{{\mathbb{R}}^{M}})}+\Vert {\dot{\nu}}\Vert _{L^{\infty
}((t_{0},t_{i}),{{\mathbb{R}}^{M}})}\right) |\overline{\nu }|.
\end{eqnarray}%
Furthermore, we use (A2) and the fact that $\partial _{\nu }e$, $\partial
_{N}e$ have linear growth to bound the right hand side of \eqref{uguagl} as
follows 
\begin{eqnarray}
\lefteqn{\langle \Psi ((\nu _{\alpha }(t_{i+1}))_{\alpha \in \mathcal{T}%
\setminus \partial \mathcal{B}_{\nu }}),\overline{\nu }\rangle }  \notag
\label{e4} \\
&\leq &cT_{\Theta }\left( 1+\Vert \dot{u}\Vert _{L^{\infty }\left(
(t_{0},t_{i-1}),{{\mathbb{R}}^{D}}\right) }+\Vert \dot{\nu}\Vert _{L^{\infty
}\left( (t_{0},t_{i-1}),{{\mathbb{R}}^{M}}\right) }\right) |\overline{\nu }|.
\end{eqnarray}%
By collecting \eqref{e1}-\eqref{e4} it follows that 
\begin{equation}
\gamma |\overline{\nu }|\leq cT_{\Theta }\left( 1+\Vert \dot{\nu}\Vert
_{L^{\infty }\left( (t_{0},t_{i}),{{\mathbb{R}}^{M}}\right) }+\Vert \dot{u}%
\Vert _{L^{\infty }\left( (t_{0},t_{i-1}),{{\mathbb{R}}^{D}}\right) }\right)
,  \label{stimaBV2s}
\end{equation}%
which in turn yields 
\begin{eqnarray*}
\lefteqn{(\gamma -cT_{\Theta })\Vert \dot{\nu}\Vert _{L^{\infty }\left(
(t_{0},t_{i}),\mathbb R^{M}\right) }} \\
&\leq &c\gamma \Vert \dot{\nu}\Vert _{L^{\infty }\left( (t_{0},t_{i-1}),{{%
\mathbb{R}}^{M}}\right) }+cT_{\Theta }\left( 1+\Vert \dot{u}\Vert
_{L^{\infty }\left( (t_{0},t_{i-1}),{{\mathbb{R}}^{D}}\right) }\right) .
\end{eqnarray*}%
Thus, there exists a constant $T_{1}=T_{1}(\gamma ,\Xi ,\eta ,\lambda
,\Lambda ,\mathcal{T})>0$ such that for $T_{\Theta }\in (0,T_{1})$ 
\begin{equation}
\left\Vert \dot{\nu}\right\Vert _{L^{\infty }\left( (t_{0},t_{i}),{{\mathbb{R%
}}^{M}}\right) }\leq c\Vert \dot{\nu}\Vert _{L^{\infty }\left(
(t_{0},t_{i-1}),{{\mathbb{R}}^{M}}\right) }+cT_{\Theta }\left( 1+\Vert \dot{u%
}\Vert _{L^{\infty }\left( (t_{0},t_{i-1}),{{\mathbb{R}}^{D}}\right)
}\right) .  \label{stima4s}
\end{equation}%
In particular, by setting $\beta _{l}:=\Vert \dot{u}\Vert _{L^{\infty
}\left( (t_{0},t_{l}),{{\mathbb{R}}^{D}}\right) }+\Vert \dot{\nu}\Vert
_{L^{\infty }\left( (t_{0},t_{l}),{{\mathbb{R}}^{M}}\right) }$, from %
\eqref{stimaLinf2s} and \eqref{stima4s} we infer 
\begin{equation*}
\beta _{i}\leq cT_{\Theta }+c\left( 1+T_{\Theta }\right) \beta _{i-1}.
\end{equation*}%
By iterating this inequality and arguing as in the proof of Proposition~\ref%
{stime}, one can deduce the $L^{\infty }$ estimate \eqref{Prop1s}.

Eventually, in order to prove the $BV$ estimate \eqref{stimaBVs} we can
reason as in the proof of Proposition~\ref{stime} by taking into account %
\eqref{Prop1s}, \eqref{stimaLinf1s} and \eqref{stimaBV2s}.
\end{proof}

\begin{remark}
\label{notorthon} The orthogonality condition of the family of shape
functions $\{\mathcal{N}_{a}\}_{a\in \mathcal{T}}$ has been imposed only for
convenience. Indeed, the same model of Section~\ref{AVIs}, in which the
potential is decomposed element-by-element, can be developed for the
synchronous setting, too.
\end{remark}

\begin{acknowledgement}
The Centre for Mathematical Research "E. De Giorgi" of the Scuola Normale
Superiore at Pisa is acknowledged for providing us an appropriate
environment for fruitful scientific interactions. PMM acknowledges also the
support of the Italian National Group for Mathematical Physics (GNFM-INDAM).
\end{acknowledgement}

\end{document}